\documentclass[%
 reprint,
superscriptaddress,
 amsmath,amssymb,
prb,
]{revtex4-2}

\usepackage{graphicx}
\usepackage{dcolumn}
\usepackage{bm}
\usepackage{hyperref}
\hypersetup{
	colorlinks=true,
	linkcolor=blue,
	citecolor=blue,
	filecolor=blue,      
	urlcolor=blue,
}
\usepackage{hypcap}
\newcommand{\commented}[1]{}
\usepackage[dvipsnames]{xcolor} 
\newcommand{\review}[1]{{\leavevmode\color{black}#1}}

\begin{document}

\preprint{APS/123-QED}


%
%
\title{\review{Unveiling the impact of temperature on magnon diffuse scattering detection in the transmission electron microscope}}

%
%
\author{Jos\'e \'Angel Castellanos-Reyes}
\email{angel.castellanos.research@gmail.com}
\affiliation{Department of Physics and Astronomy, Uppsala University, Box 516, 75120 Uppsala, Sweden}

\author{Paul Zeiger}%
\affiliation{Department of Physics and Astronomy, Uppsala University, Box 516, 75120 Uppsala, Sweden}
\author{Anders Bergman}
\affiliation{Department of Physics and Astronomy, Uppsala University, Box 516, 75120 Uppsala, Sweden}
\author{Demie Kepaptsoglou}
\affiliation{SuperSTEM Laboratory, SciTech Daresbury Campus, Daresbury WA4 4AD, United Kingdom}
\affiliation{Department of Physics, University of York, York YO10 5DD, United Kingdom}
\author{Quentin M. Ramasse}
\affiliation{SuperSTEM Laboratory, SciTech Daresbury Campus, Daresbury WA4 4AD, United Kingdom}
\affiliation{School of Chemical and Process Engineering, University of Leeds, Leeds LS2 9JT, United Kingdom}
\affiliation{School of Physics and Astronomy, University of Leeds, Leeds LS2 9JT, United Kingdom}
\author{Juan Carlos Idrobo}

\affiliation{Materials Science and Engineering Department, University of Washington, Seattle, Washington 98195, USA}
\author{J\'an Rusz}
\email{jan.rusz@physics.uu.se}
\affiliation{Department of Physics and Astronomy, Uppsala University, Box 516, 75120 Uppsala, Sweden}

\date{\today}
%
%
\begin{abstract}
Magnon diffuse scattering (MDS) signals could, in principle, be studied with high spatial resolution in scanning transmission electron microscopy (STEM), thanks to recent technological progress in electron energy loss spectroscopy. However, detecting MDS signals in STEM is technically challenging due to their overlap with the much stronger thermal diffuse scattering (TDS) signals. In bcc Fe at 300~K, MDS signals greater than or comparable to TDS signals \review{have been predicted to} occur under the central Bragg disk, well into a currently inaccessible energy-loss region. Therefore, to successfully detect MDS in STEM, it is necessary to identify conditions in which TDS and MDS signals can be distinguished from one another in regions outside the central Bragg disk. Temperature may be a key factor due to the distinct thermal signatures of magnon and phonon signals. In this work, we present a study on the effects of temperature on MDS and TDS in bcc Fe---considering a detector outside the central Bragg disk \review{and a fixed convergent electron probe}---using the frozen phonon and frozen magnon multislice methods. Our study reveals that neglecting the effects of atomic vibrations causes the MDS signal to grow approximately linearly up to the Curie temperature of Fe, after which it exhibits less variation. The MDS signal displays an alternating behavior due to dynamical diffraction, instead of increasing monotonically as a function of thickness. The inclusion of the \review{effects of atomic vibrations through a complex atomic electrostatic potential} causes the linear growth of the MDS signal to change to a \review{non-linear} behavior that exhibits a predominant peak \review{for a sample of thickness 16.072~nm at 1100~K}. In contrast, the TDS signal grows more linearly than the MDS signal \review{through the studied temperature range} but still exhibits appreciable dynamical diffraction effects. An analysis of the signal-to-noise ratio (SNR) shows that the MDS signal can be a statistically significant contribution to the total scattering intensity under \review{realizable} measurement conditions and \review{feasible} acquisition times. For example, our study found that a SNR of 3 can be achieved with a beam current of 1~nA in less than 30 minutes \review{for the 16.072~nm thick bcc Fe sample at 1100~K}.
\end{abstract}
\maketitle

%
%
\section{\label{sec:introduction}Introduction}

Scanning transmission electron microscopy (STEM) is a powerful and versatile technique to study and characterize micro- and nanostructures~\cite{carter2009transmission}. Recent progress in STEM monochromators and spectrometers has made it possible to perform electron energy loss spectroscopy (EELS) with sub-10 meV energy resolution at nanometric and atomic spatial resolutions~\cite{krivanek2014nature, krivanek2019ultramicroscopy, dellby2020ultra, dellby2022multi}. This has opened the possibility for high-spatial-resolution STEM-EELS studies of elementary excitations in the zero-to-few-hundreds meV range, such as molecular vibrations, infrared plasmons, and phonons~\cite{lagos2022advances}. It has been pointed out that \review{high-spatial-resolution} STEM-EELS could, in principle, be performed also for magnons~\cite{Lyon2021PRB, mendis2021, mendis2022}, since their excitation energies lie in the same range~\cite{Eriksson2017}. 

Magnons are quanta of collective spin excitations (quantized spin waves), pictured semiclassically as waves of precessing magnetic dipole moments~\cite{Eriksson2017}. These quasiparticles lie at the core of the current understanding of the ordered magnetism of solids~\cite{kittel2018introduction, mohn2006magnetism}. Therefore, studying magnons at high spatial resolution in STEM would be relevant not only for magnetic solid-state technologies (such as spintronics, spin caloritronics~\cite{zutic2004,pulizzi2012spintronics}, and magnonics~\cite{Barman2021}) but also for the foundations of solid-state magnetism.

\review{Other inelastic-scattering techniques have been already employed to study magnons~\cite{Eriksson2017}. In particular, magnons have been probed with energy and momentum resolution using reflection EELS (REELS) and spin-polarised EELS (SPEELS)~\cite{VOLLMER20042126,Zakeri2013,ibach2017electron}. However,  both SPEELS and REELS setups are limited to surface or thin-film studies due to the low energy and limited penetration depth of the electron probe. Furthermore, these techniques cannot achieve the spatial resolution offered by STEM.}

Detecting magnon signals in STEM is technically challenging since they are typically orders of magnitude less intense than the so-called thermal diffuse scattering (TDS) signals~\cite{Lyon2021PRB, mendis2022}---produced by the inelastic scattering of the electron probe due to lattice vibrations (i.e., phonons). For example, in Ref.~\cite{Lyon2021PRB} it was reported that the simulated TDS signal for bcc Fe at 300~K is four orders of magnitude greater than the corresponding magnon diffuse scattering (MDS) signal. \review{Furthermore, simulations in Ref.~\cite{mendis2022} of the same system predicted another challenge: that MDS signals exceeding or matching TDS signals are observed solely for scattering angles $\alt  0.5$~mrad (refer to Fig.~2(b) in Ref.~\cite{mendis2022} along with the associated discussion).} This region corresponds---through the dispersion relation---to magnons with energies below 10~meV~\cite{Eriksson2017}, practically on the current energy resolution limit of monochromated EELS~\cite{krivanek2014nature, krivanek2019ultramicroscopy, dellby2020ultra, dellby2022multi}.

Hence, to achieve MDS detection in STEM at high-spatial resolution, it is necessary to find conditions in which MDS and TDS signals can be told apart. In particular, as has been argued in Ref.~\cite{Lyon2021PRB}, temperature could play a decisive role for this purpose\review{, especially since experiments featuring STEM, equipped with precise temperature control capabilities, can already be conducted successfully~\cite{IdroboPhysRevLett.120.095901,lagos2018thermometry,KikkawaPhysRevB.106.195431,wehmeyer}. }

In this Article, we investigate the behavior of MDS at different temperatures and explore the possibility of temperature-aided detection of MDS in STEM.
Employing the prototypical bcc Fe as the magnetic system and the methodology developed in Ref.~\cite{Lyon2021PRB}, we investigate the temperature dependence of simulated MDS signals 
\review{considering a fixed convergent electron beam.}
%
From the studied cases, we establish optimal combinations of temperature and sample thickness having the highest MDS signals.
In particular, we focus on signals surrounding the central Bragg disk to explore and address the challenge reported in Ref.~\cite{mendis2022}. Finally, we compare our results with TDS simulations and discuss the feasibility of MDS detection in STEM. 
%
\review{Our study, not incorporating energy resolution, represents a worst-case scenario for magnon detection in STEM. Additionally, it's worth noting that we do not include the use of a scanning probe. Thus, many of the results presented below are relevant for both STEM and conventional transmission electron microscopy (TEM).}


%
%
\section{\label{sec:methods}Methods}

To simulate the inelastic electron-probe scattering on a specimen at a certain temperature in \review{(S)TEM---we will use the term (S)TEM when referring to both TEM and STEM---}, it is necessary to have a model for the specimen at the considered temperature and to implement a method for electron-beam propagation through it.
In this work, the inelastic signals---TDS and MDS---of ferromagnetic bcc Fe are obtained following the methodology of Ref.~\cite{Lyon2021PRB}. Explicitly, the TDS signals are calculated via the frozen phonon multislice (FPMS) method~\cite{Loane1991} and the MDS signals via the analogous frozen magnon multislice (FMMS) method, originally introduced in Ref.~\cite{Lyon2021PRB}. These methods are named ``frozen'' because each electron from the \review{(S)TEM} probe travels with a relativistic speed, interacting with the specimen in a time on the order of tens of attoseconds, at which the motion of atoms and their \review{magnetic moments look practically ``frozen.''}

To simulate the electron-beam propagation through the specimen in FPMS, the conventional multislice method~\cite{kirkland2010} is employed. In FMMS, to account for the effects of spins and magnetism, the Pauli multislice method\review{---a multislice approach to solving the relativistically-corrected paraxial Pauli equation~\cite{Edstrom2016PRB,Edstrom2016PRL}---}is utilized.
We employed an in-house developed software for both multislice methods.

For TDS calculations at a given temperature, the magnetic moments of Fe atoms are completely ignored, and the dynamics of the Fe atomic vibrations (phonons) are obtained through molecular dynamics (MD) simulations. 
Meanwhile, for the MDS calculations, atomistic spin dynamics (ASD)~\cite{Eriksson2017} simulations are employed to obtain the dynamics of the magnons---i.e. the evolution of the precessing magnetic moments of Fe atoms (of imposed constant magnitude)---assuming that the atomic positions are kept fixed.
ASD simulations accurately model the dynamics of thermally excited \review{magnetic moment configurations} in a manner analogous to how MD does for atomic vibrations.
In FPMS, the TDS signal is obtained by sampling over the possible atomic displacements configurations~\cite{Loane1991,Zeiger2020}. Analogously, the MDS signal in FMMS is computed by sampling over the \review{magnetic moment configurations}~\cite{Lyon2021PRB}. In both FPMS and FMMS, the inelastic signal at the diffraction plane, for a given temperature, is calculated as the difference between the so-called \textit{incoherent} and \textit{coherent} intensities~\cite{VanDyck2009,Forbes2010QEP,Lyon2021PRB}. On the one hand, the incoherent intensity---corresponding to the total scattered signal $I_\text{tot}$ in the diffraction plane---is the average, over all samples, of the exit wavefunctions' intensities (squared amplitudes). On the other hand, the coherent intensity is the squared amplitude of the averaged exit wavefunctions, and it corresponds to the purely elastic scattering signal $I_\text{ela}$ in the diffraction plane. Therefore, the inelastic signal $I_\text{ine}(T)$ at temperature $T$ is given by 
%
%
\begin{equation}
I_\text{ine}(T)
=
I_\text{tot}(T)
-
I_\text{ela}(T),
\label{Eq:Iinelastic}
\end{equation}
%
where ``ine'' stands for MDS in the case of FMMS, and TDS in the case of FPMS. \review{It is worth noting that, in practice, additional inelastic signals, such as valence and core losses, as well as plasmon scattering, coexist alongside TDS and MDS~\cite{carter2009transmission}.  Due to their different range of energy losses, these additional signals can be effectively filtered from experimental diffraction patterns and, consequently, will not be further considered in this work.}

We have chosen ferromagnetic bcc Fe as our model system because it is a prototypical magnetic material for which magnons have been detected using electron beams~\cite{Zakeri2013}. Moreover, the methodology discussed above has already been tested in Ref.~\cite{Lyon2021PRB} for bcc Fe. 

\begin{figure}[h]
	\centering
	\includegraphics[width=0.5\linewidth]{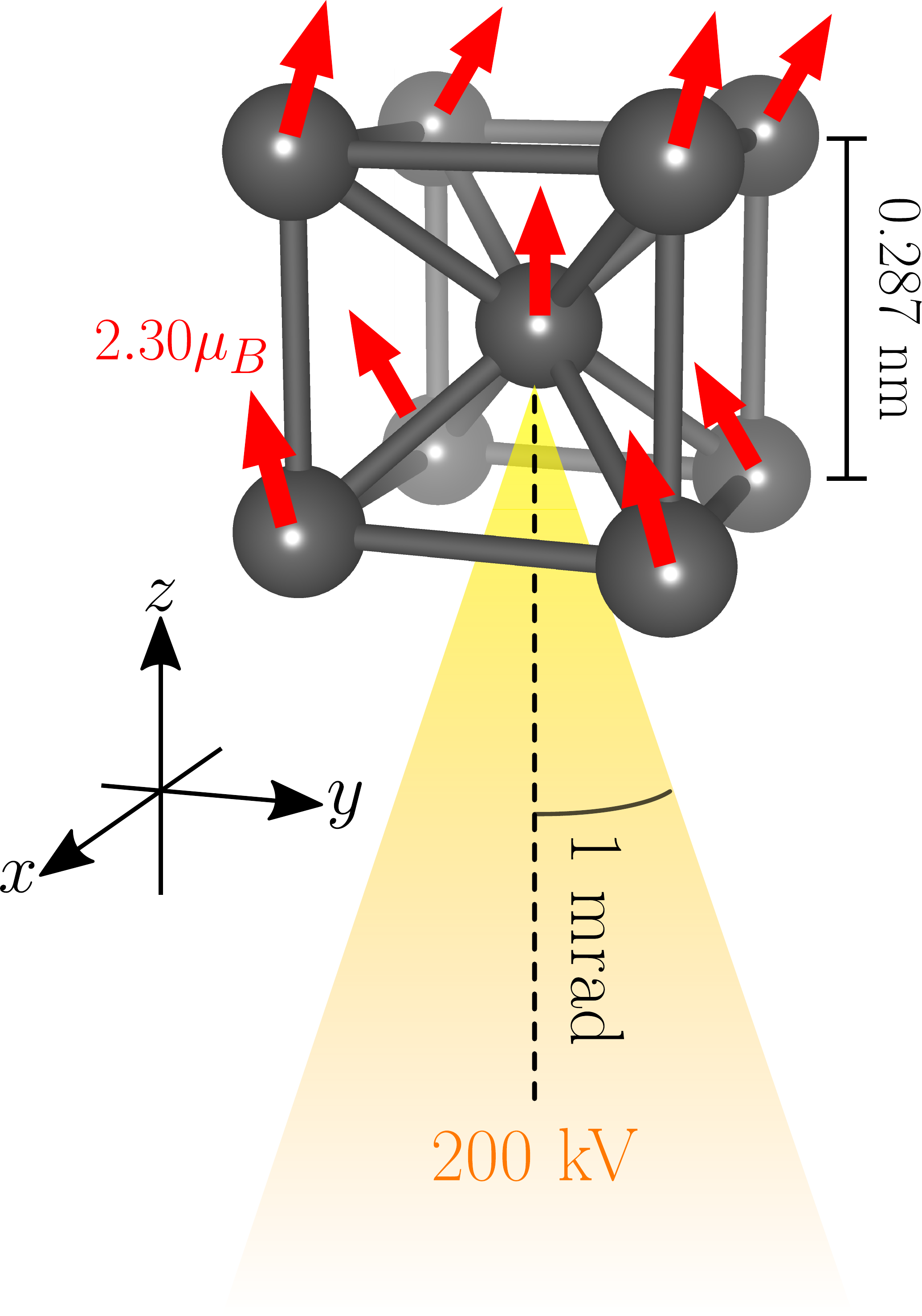}
	\caption{Scheme of the system under consideration \review{(not-to-scale)}. A 200~kV aberration-free \review{(S)TEM} electron probe, propagating in $z$ direction with 1~mrad convergence semiangle, \review{illuminates a sample of bcc Fe---having $z$ as its [001] direction---at a certain temperature}, with lattice parameter 0.287~nm and 2.30~$\mu_B$ atomic magnetic moment ($\mu_B$ is the Bohr magneton).}

	\label{Fig:system}
\end{figure}

For the calculations, we have employed supercells $\mathcal{S}_t$ consisting of $20 \times 20 \times (14t)$ repetitions of the bcc Fe unit cell (in $x,y,$ and $z$ directions, respectively; see Fig.~\ref{Fig:system}), with $t \in \{ 1,2,3,4, 5 \}$ to account for five different thicknesses, of dimensions $5.74 \times 5.74 \times (4.018t)$~nm$^3$, containing $11200t$ atoms. Periodic boundary conditions were considered in $x$ and $y$ directions \review{(and in $z$ direction for the $t=5$ supercells)}.
%
%

To implement FMMS, we generated a representative sampling of the \review{magnetic moment configurations} from ASD simulations using the \texttt{UppASD} code~\cite{UppASD,Eriksson2017}~\footnote{We have used \texttt{UppASD} version 5.0.}. We considered the Heisenberg Hamiltonian with exchange interactions and magnetic moments computed ab initio with the scalar-relativistic \texttt{SPRKKR} code~\cite{Ebert2011}. 
The magnitude of Fe magnetic moments was 2.30~$\mu_B$ (where $\mu_B$ is the Bohr magneton).
To account for the effect of the microscope's objective lens, we have included a 1 T external static magnetic field oriented along the positive $z$ direction.
A sample of 101 configurations per temperature--- from 0~K to 1700~K---was obtained by taking a \textit{snapshot} (i.e., a static configuration), every 10~fs, out of an ASD simulation with a 0.1~fs time step. To minimize the correlation between different snapshots, we set a large Gilbert damping parameter $\alpha=0.5$ in the simulations. Then, for each snapshot, we performed a Pauli multislice simulation. Finally, the MDS signal at a given temperature was obtained using the coherent and incoherent averages (over all the snapshots) as in Eq.~(\ref{Eq:Iinelastic}). A discussion about the level of convergence of our calculations in terms of the accuracy of the computed averages is presented in Appendix~\ref{app:statistical}.

Analogously, we performed FPMS simulations using snapshots sampled from trajectories of constant temperature MD simulations (``$NVT$ ensemble'') in the \texttt{LAMMPS} software \cite{lammps_web,LAMMPS_paper_2022}\footnote{We have used \texttt{LAMMPS} version 23 Jun 2022 -- Update 1.}. The size of the supercell was set in the same way as in the \texttt{UppASD} calculations described above. The simulations were run using a Nosé-Hoover thermostat, which maintained the specified temperature with a temperature damping parameter $T_{\mathrm{damp}} = 100$~fs. In order to account for thermal expansion, the average lattice parameter in the $NVT$ ensemble simulations was determined from constant temperature and constant pressure MD simulations (``$NPT$ ensemble'') for each temperature. The time step of the MD simulation was set to 1~fs and the interatomic forces between Fe atoms were described by a so-called \emph{embedded-atom method} (EAM) potential \cite{Mendelev2003PhilMag}. Similar to the case of the FMMS simulations, we sample 101 configurations per temperature from the MD trajectories in the $NVT$ ensemble by taking a snapshot of the \emph{atomic positions} every 1000~fs after an initial thermal equilibration time of 10000~fs.

In the conventional and Pauli multislice simulations, following the discussion of Ref.~\cite{Lyon2021PRB} regarding the resolution of inelastic signals in the diffraction plane, we have employed a \review{fixed} 200~kV aberration-free electron probe focused on the entrance surface of the supercell, with a 1~mrad convergence semiangle, \review{propagating in [001] direction (corresponding to $z$ direction). This is illustrated in Fig.~\ref{Fig:system}, not drawn to scale~\footnote{The smallest achievable size of the employed electron probe on the specimen is $d_0=1.5311$~nm, corresponding roughly to a linear extension of 5.3 bcc Fe unit cells. This value can be computed from the diffraction limit $d_0=0.61\lambda/\alpha_0$~\cite{pennycook2011scanning}, using the probe's convergence semiangle $\alpha_0=1$~mrad and its 200~kV de-Broglie wavelength $\lambda=0.00251$~nm~\cite{kirkland2010}.}}. For the supercell $\mathcal{S}_t$, the multislice calculations were performed on a regular grid $\mathcal{G}_t$ consisting of $1000 \times 1000 \times (420t)$ points in $x,y,$ and $z$ directions, spanning the entire supercell.

The magnetic field $\mathbf{B}(\mathbf{r})$ and vector potential $\mathbf{A}(\mathbf{r})$\review{, at position $\mathbf{r}$,} produced by Fe magnetic moments on a given snapshot (used for multislice simulations) were calculated using the parametrization by Lyon and Rusz~\cite{Lyon2021AC}, which has been successfully benchmarked against density functional theory calculations of bcc Fe. 

\review{
For the electrostatic potential $V(\mathbf{r})$ of Fe atoms, we employed the parametrization developed by Peng et al.~\cite{DUDAREV199586,Peng1996}, which incorporates absorption effects due to phonons by considering a complex $V(\mathbf{r})$ (see Appendix~\ref{app:meansquaredisplacements} for details). In particular, this complex potential incorporates the Debye-Waller factor (DWF)~\cite{kittel2018introduction}\review{---leading to weaker elastic scattering at high angles and stronger at low angles---}in both its real part (elastic potential) and in its imaginary part (absorptive potential). Hence, in our FPMS simulations, we exclusively consider the elastic potential without DWF, since the effect of atomic vibrations is already fully included through averaging over snapshots. Conversely, our FMMS simulations incorporate the complete complex potential to model (in first approximation) the absorption by phonons and the corresponding attenuation of elastic signals due to thermal motions. The mean-squared displacements used for the implementation of the DWF were computed from a running average, computed at every time step in the aforementioned MD simulations in the $NVT$ ensemble, and are presented in Table~\ref{tab:meansquaredisplacements} in Appendix~\ref{app:meansquaredisplacements}.
}

In all cases, $V(\mathbf{r})$, $\mathbf{B}(\mathbf{r})$, and $\mathbf{A}(\mathbf{r})$ were computed in the gridpoints of $\mathcal{G}_t$ surrounding each Fe atom up to a specified cutoff distance $r_\text{cut}$, \review{beyond which} all are set to zero. The specific value of $r_\text{cut}$ used in each case was chosen as a compromise between numerical accuracy and computational resources demand, see Appendix~\ref{app:cutoff}.


%
%
\section{\label{sec:simulations}Effects of the temperature on MDS and feasibility of temperature-aided detection}

The aim of this work is to investigate the behavior of MDS at different temperatures, particularly, to explore the possibility of temperature-aided MDS detection in \review{(S)TEM}. Therefore, we start our study in Subsection~\ref{sec:diffandADF} presenting the general features of the resulting electron-probe diffraction patterns. At this first stage, we select a detector that collects relevant MDS signals surrounding the central Bragg disk and study these signals as a function of temperature in the following. 

In Subsection~\ref{sec:nodwfadf} we first study the MDS while completely ignoring the effects of atomic vibrations\review{, considering only the elastic potential without the Debye-Waller factor (DWF) [i.e. setting $B=0$ and $U_n^\text{(abs)}=0$ in Eqs.~(\ref{Eq:PengunRe}) and (\ref{Eq:Penguntds})]}. Therein, we present the results of the MDS as a function of temperature for all the specimen thicknesses considered.

\review{
Two significant effects of atomic vibrations in (S)TEM are the attenuation of elastic signals with increasing scattering angle and/or temperature and temperature-dependent absorption due to phonons~\cite{kittel2018introduction, carter2009transmission,Peng1996,DUDAREV199586}. These effects can be incorporated into static-lattice calculations by considering the complete complex electrostatic potential, which includes the Debye-Waller factor (DWF) in both the elastic and absorptive potentials. Hence, to continue our investigations, in Subsection~\ref{sec:dwf} we examine how these effects alter the MDS signal with varying temperatures, representing an initial approximation to the incorporation of atomic vibrations.

%

To complement this, in Subsection~\ref{sec:tds}, we present simulations of the TDS signal, where we completely ignore the magnetic moments of Fe and consider only the elastic electrostatic potential without the DWF. We then compare and contrast these results with the MDS signal. Lastly, in Subsection~\ref{sec:eels}, we delve into the implications of our findings for the successful detection of MDS in (S)TEM.
%
}

All the intensities presented in the figures of this work are divided by the total intensity of the incident electron beam integrated over the whole diffraction plane, $I_0$, to show dimensionless results. Moreover, when plotted in regions of the diffraction plane, as a function of the scattering angle, they actually correspond to intensities integrated over pixels. This is the case for Figs.~\ref{Fig:nodwfdiff}, \ref{Fig:tds}, \ref{Fig:statisticalMDS}, and \ref{Fig:cutoff}. The size of a pixel in our calculations (that can be computed from the parameters described in Section~\ref{sec:methods}) is 0.19~mrad$^2$.


%
%
%
\subsection{\label{sec:diffandADF}MDS diffraction patterns and selection of an annular dark-field detector}

\begin{figure}[h]
	\centering
	\includegraphics[width=\linewidth]{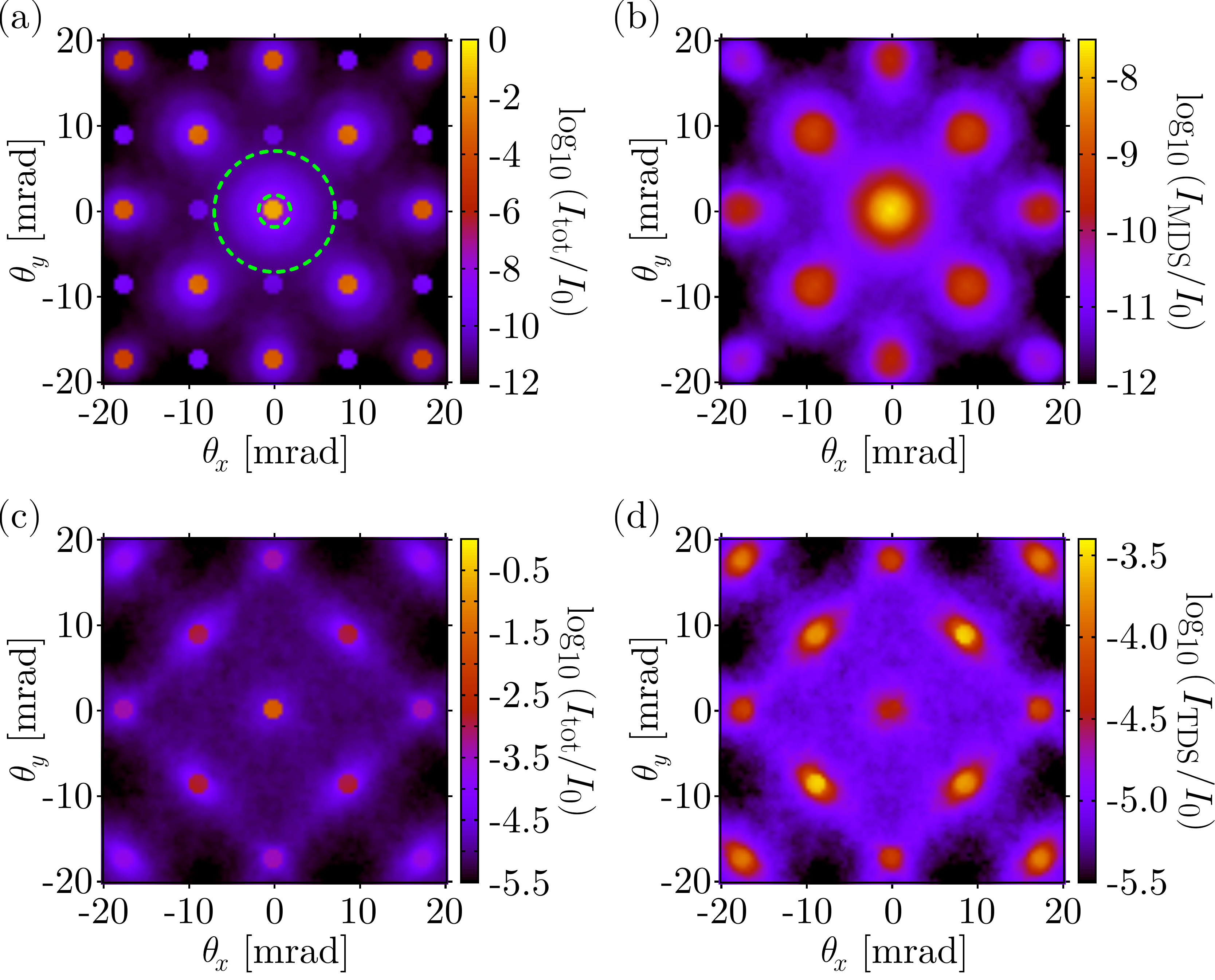}
	\caption{\review{Diffraction patterns computed for bcc Fe with a thickness of 16.072~nm at 1100~K, considering the full complex potential $V(\mathbf{r})$. (a) FMMS total signal $\log_{10}(I_\text{tot}/I_0)$ [$I_0$ denotes the total intensity of the incident electron beam integrated over the whole diffraction plane] and annular dark-field detector (ADF; shown in green dashed lines). (b) Magnon diffuse scattering (MDS) signal $\log_{10}(I_\text{MDS}/I_0)$. For reference, the lower panels show FPMS diffraction patterns showing the TDS---panel (c) corresponds to (a), and panel (d) corresponds to (b).}
 }
	\label{Fig:nodwfdiff}
\end{figure}

\review{
We simulated MDS signals for different temperatures and thicknesses while keeping the atomic positions fixed,
both when considering only the elastic electrostatic potential while ignoring the DWF, and when considering the full complex potential $V(\mathbf{r})$.
%
%
The relevant features of all the resulting diffraction patterns can be appreciated in \review{the upper row of} Fig.~\ref{Fig:nodwfdiff}, showing results for bcc Fe of 16.072~nm thickness at 1100~K, including the full complex $V(\mathbf{r})$. Additionally, we included the corresponding TDS results in the lower row of Fig.~\ref{Fig:nodwfdiff} as a reference.
}
%
%
%

The general behavior of the total signal is illustrated in Fig.~\ref{Fig:nodwfdiff}(a), showing $\log_{10}(I_\text{tot}/I_0)$. The total signal consists of Bragg disks, alternating on high and low \review{(kinematically forbidden reflections)} intensities, with low-intensity lobes surrounding the high-intensity Bragg disks.

The MDS signal, $\log_{10}(I_\text{MDS}/I_0)$, computed from Eq.~(\ref{Eq:Iinelastic}), is shown in Fig.~\ref{Fig:nodwfdiff}(b). In particular, it can be appreciated that the MDS signal is concentrated around the high-intensity Bragg disks, vanishing away from the center of the diffraction plane. We employed $r_\text{cut} = 1$~nm in Fig.~\ref{Fig:nodwfdiff} for a better resolution of the MDS near the Bragg disks (see Appendix~\ref{app:cutoff}). In particular, the highest MDS signal is located within the central Bragg disk, in agreement with Ref.~\cite{mendis2022}. Therefore, for experimental detection, it is relevant to analyze the MDS signal at small scattering angles surrounding the central Bragg spot. 

Thus, to study the effects of temperature, we considered an annular dark-field (ADF) detector~\cite{kirkland2010} of inner collection semiangle 2~mrad and outer collection semiangle 7~mrad [illustrated by the green dashed lines in Fig.~\ref{Fig:nodwfdiff}(a)]---to avoid all Bragg disks, including the \review{kinematically forbidden reflections}. Hence, this detector collects only the MDS signal. 
In particular, as discussed in Appendix~\ref{app:cutoff}, the calculations in this ADF detector are already converged at $r_\text{cut} = 0.4$~nm---having MDS signals two orders of magnitude greater than the error coming from the averaging process (see Appendix~\ref{app:statistical}).
Therefore, in the following, we study the effects of temperature on the signals collected by the aforementioned ADF detector using $r_\text{cut} = 0.4$~nm.


    %
    %
\subsection{\label{sec:nodwfadf}MDS neglecting the effect of the atomic vibrations}

\begin{figure}[h]
	\centering
	\includegraphics[width=\linewidth]{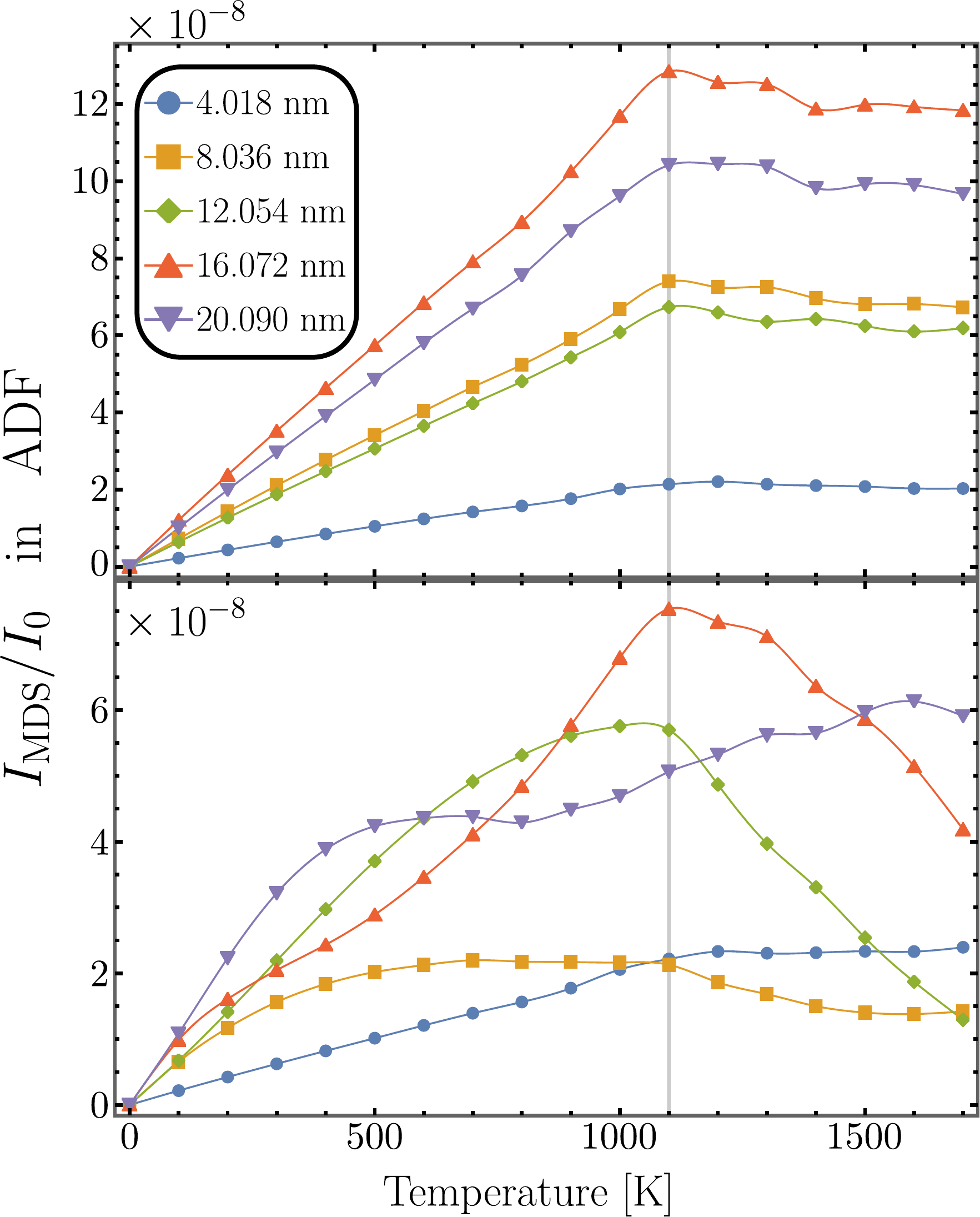}
	\caption{MDS signals collected by the selected ADF detector for bcc Fe of different thicknesses (indicated in the legend of the top panel) as a function of temperature. The top panel shows $I_\text{MDS}/I_0$ neglecting the effect of the atomic vibrations, while the bottom panel  \review{incorporates these effects through the complex $V(\mathbf{r})$}. The gray vertical line indicates the Curie temperature of the samples. Solid lines joining the computed values are only a guide to the eye.
 }
	\label{Fig:adf}
\end{figure}

In the top panel of Fig.~\ref{Fig:adf}, we show the simulated MDS signals as a function of temperature for the five different thicknesses considered in this work (in all cases, the continuous lines joining the computed values are only a guide to the eye). Specifically, we present $I_\text{MDS}/I_0$ collected by the selected ADF detector up to 1700~K (the melting temperature of our system is around 1800~K). It can be appreciated in this panel that, for all thicknesses, the MDS signal grows approximately linearly up to $\approx 1100$~K, corresponding \review{roughly} to the Curie temperature ($T_\text{C}$) of \review{the} samples. Above $T_\text{C}$, the linear increment stops, giving place to a less-varying behavior. These features are consistent with the semiclassical picture of the interaction, in which the MDS signal would increase with the randomness in the orientation of the magnetic moments (this randomness reaches its maximum for $T\geq T_\text{C}$). 
%
%
%

It could be expected that the MDS signal increases with the thickness. However, in the top panel of Fig.~\ref{Fig:adf}, the signal corresponding to the thickness $8.036$~nm is greater than the one of $12.054$~nm. Also, the results of $16.072$~nm are greater than those of $20.090$~nm. This alternating behavior is due to dynamical diffraction (multiple scattering)~\cite{carter2009transmission}. In particular, increasing the thickness of the specimen for the electron beam propagation can lead to constructive and destructive interference conditions for the coherent intensity that are \review{negligible} at the lowest thickness.



%
%
\subsection{\label{sec:dwf}Effects of the \review{atomic vibrations} on MDS}

\review{
The effects of including the atomic vibrations, as a first approximation, by fully incorporating the complex potential $V(\mathbf{r})$ into the MDS calculations can be appreciated in the lower panel of Fig.~\ref{Fig:adf}. This panel displays the signals collected by the selected ADF detector, following the same format as the top panel within the same figure.
}
%

For the thinnest specimen considered (4.018~nm), the ADF-collected signal in the bottom panel of Fig.~\ref{Fig:adf} again grows linearly up to $T_\text{C}$, saturating at a slightly higher value than in the top panel. However, the higher sample thicknesses display a 
qualitatively different temperature dependence, in which the saturation and linearity disappear. \review{Instead, there is a \review{contrasting} non-linear behavior, which presents a predominant peak for the sample of thickness 16.072~nm at 1100~K.}
%
%

The changes between the behavior of the MDS signals in the top and bottom panels of Fig.~\ref{Fig:adf} come from the fact that the DWF \review{and the absorption} vary with temperature---see Appendix~\ref{app:meansquaredisplacements}. \review{In particular}, an effect of the DWF is to reduce the probability that the electron probe scatters to higher angles. This reduction becomes stronger as the temperature increases~\cite{kittel2018introduction}. Therefore, the DWF will modify the interference effects that produced the alternating (thickness) behavior in the case of the top panel of Fig.~\ref{Fig:adf}, where there was no DWF. In general, dynamical diffraction effects~\cite{carter2009transmission}, which affect the thickness dependence of the electron scattering signals, will be modified by the DWF. 
%


%
%
\subsection{\label{sec:tds} Comparison between MDS and TDS}

\begin{figure}[h]
	\centering
	\includegraphics[width=\linewidth]{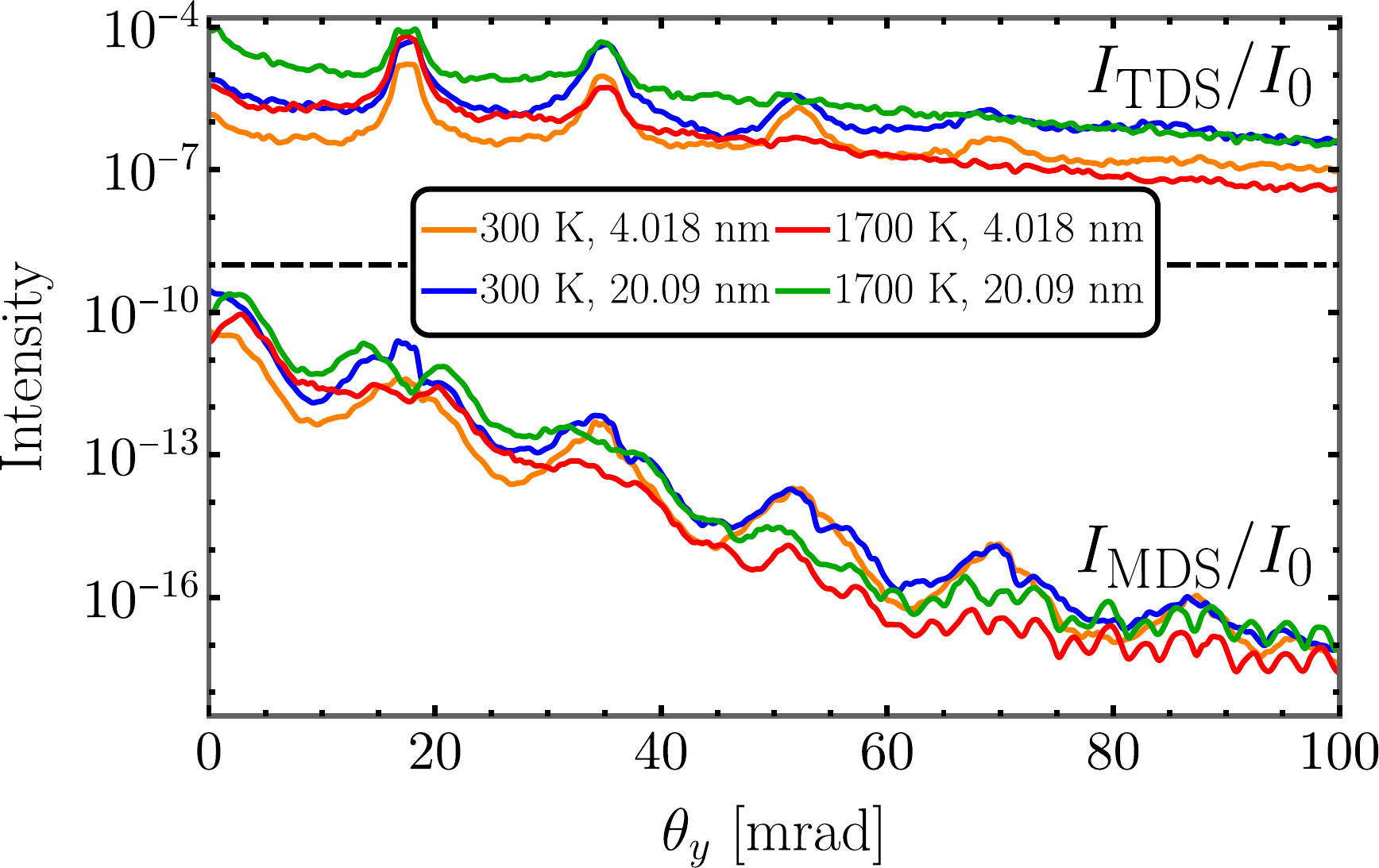}
	\caption{$I_\text{MDS}/I_0$ [including the \review{effects of atomic vibrations through the complex potential $V(\mathbf{r})$}] and $I_\text{TDS}/I_0$ as a function of the scattering angle $\theta_y$, with $\theta_x=0$, for bcc Fe---at 300 K and thickness 4.018~nm (orange curves), at 300 K and thickness 20.09~nm (blue curves), at 1700 K and thickness 4.018~nm (red curves), and at 1700 K and thickness 20.09~nm (green curves). In all cases, $r_\text{cut} = 0.4$~nm has been used. The TDS signals, always greater than the corresponding MDS signals, are located in the upper portion of the plot (above the horizontal black dashed line), while the MDS ones are in the lower portion.
 }
	\label{Fig:tds}
\end{figure}

In Ref. \cite{Lyon2021PRB} it was reported that the TDS signal was typically at least four orders of magnitude greater than the corresponding TDS signal at 300 K. We have found that this is also the case at the different temperatures and thicknesses considered in this work. This is illustrated in Fig.~\ref{Fig:tds} showing vertical profiles of the TDS and MDS [including \review{the complex $V(\mathbf{r})$}] signals through the center of the diffraction plane (i.e., as a function of the scattering angle $\theta_y$, with $\theta_x=0$). 

%
%

Specifically, in Fig.~\ref{Fig:tds} we show $I_\text{TDS}/I_0$ and $I_\text{MDS}/I_0$ for the thickest (blue and green curves) and the thinnest (orange and red curves) bcc Fe samples at 300 K (in orange and blue curves) and 1700 K (in red and green curves).
In all cases, $r_\text{cut} = 0.4$~nm has been used.
We have employed the same colors for the corresponding MDS and TDS signals, since they are well separated (we have included a horizontal black dashed line dividing them). In particular, it can be appreciated that the difference between the TDS and MDS signals becomes even larger at higher scattering angles. Therefore, the region of interest for MDS detection---now in the presence of the TDS signal---is again that of small scattering angles. This, together with avoiding the Bragg disks, supports the choice of the same ADF detector used for the MDS studies in the previous subsections.

\begin{figure}[h]
	\centering
	\includegraphics[width=\linewidth]{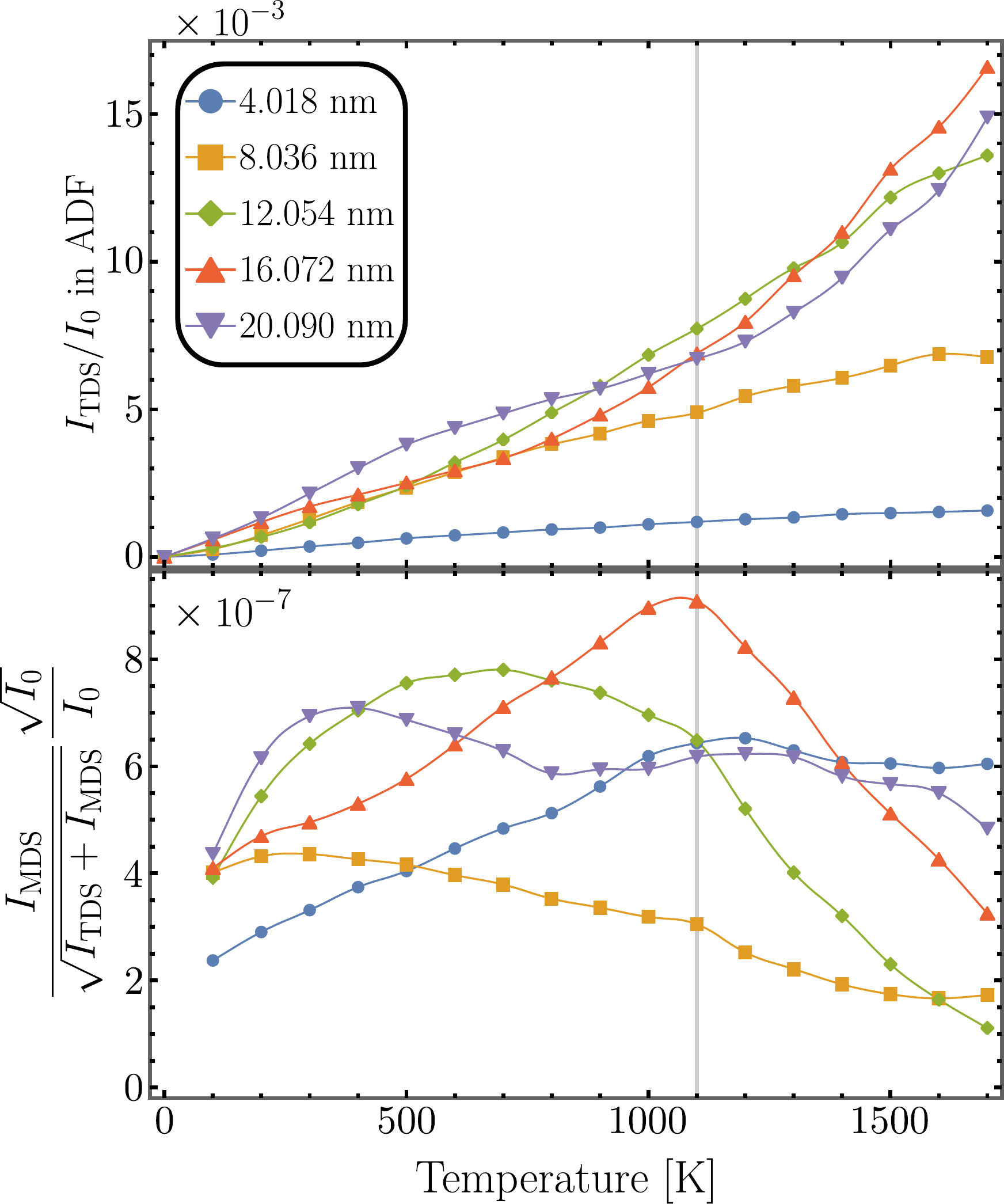}
	\caption{
 Top panel: $I_\text{TDS}/I_0$ collected by the ADF detector as a function of temperature in the same format as Fig.~\ref{Fig:adf}. Bottom panel: signal-to-noise ratio (SNR) evaluated for a dose of one electron. Line colors follow the legend of the top panel.
    }
	\label{Fig:SNRandTDS}
\end{figure}

In the top panel of Fig.~\ref{Fig:SNRandTDS}, we present the signal $I_\text{TDS}/I_0$ collected by the ADF detector, as a function of temperature, for the different sample thicknesses, in the same format as in Fig.~\ref{Fig:adf}. It can be observed that the TDS signal is about five orders of magnitude greater than the corresponding MDS signal (see Fig.~\ref{Fig:adf}). In contrast to the MDS case \review{(including the complex potential)}, the TDS curves grow more linearly with the temperature \review{(up to $T_\text{C}$ and above)}, but dynamical diffraction effects can also be appreciated. 
%
%
%

To determine the optimal combination of temperature and sample thickness for MDS detection, it is relevant to consider the signal-to-noise ratio (SNR) in the ADF detector, which is given in this case by
%
%
\begin{equation}
\text{SNR}
=
\left(
\frac
{I_\text{MDS}}
{
\sqrt{I_\text{TDS} + I_\text{MDS}}
}
\frac{\sqrt{I_0}}{I_0}
\right)
%
\sqrt{
\frac
{i_b t_a}
{e}
}
,
\label{Eq:SNR}
\end{equation}
%
where $i_b$ is the \review{(S)TEM}-electron-beam current, $t_a$ is the acquisition time, $e$ is the elementary electric charge, and the inelastic intensities are those collected by the detector. 

In the bottom panel of Fig.~\ref{Fig:SNRandTDS} we show the SNR evaluated for an electron dose of one electron [i.e., for $i_b t_a / e = 1$ in Eq.~(\ref{Eq:SNR})] for all the temperatures and sample thicknesses considered. It can be observed that the optimal detection setup corresponds to \review{the} sample of 16.072~nm thickness at \review{1100} K. This is a consequence of the contrasting behavior of the MDS and TDS signals for the 16.072~nm specimen around \review{1100} K, which can be appreciated in the top panel of Fig.~\ref{Fig:SNRandTDS} and the bottom panel of Fig.~\ref{Fig:adf}.
%
While the MDS has a peak at \review{1100}~K, with a distinctive concave behavior, the TDS presents \review{an increasing and} slightly convex behavior in the same region. Could this change be detected and resolved in current \review{(S)TEM} machines? A positive answer would imply a method for temperature-aided detection of MDS in \review{(S)TEM}.

\begin{figure}[h]
	\centering
	\includegraphics[width=\linewidth]{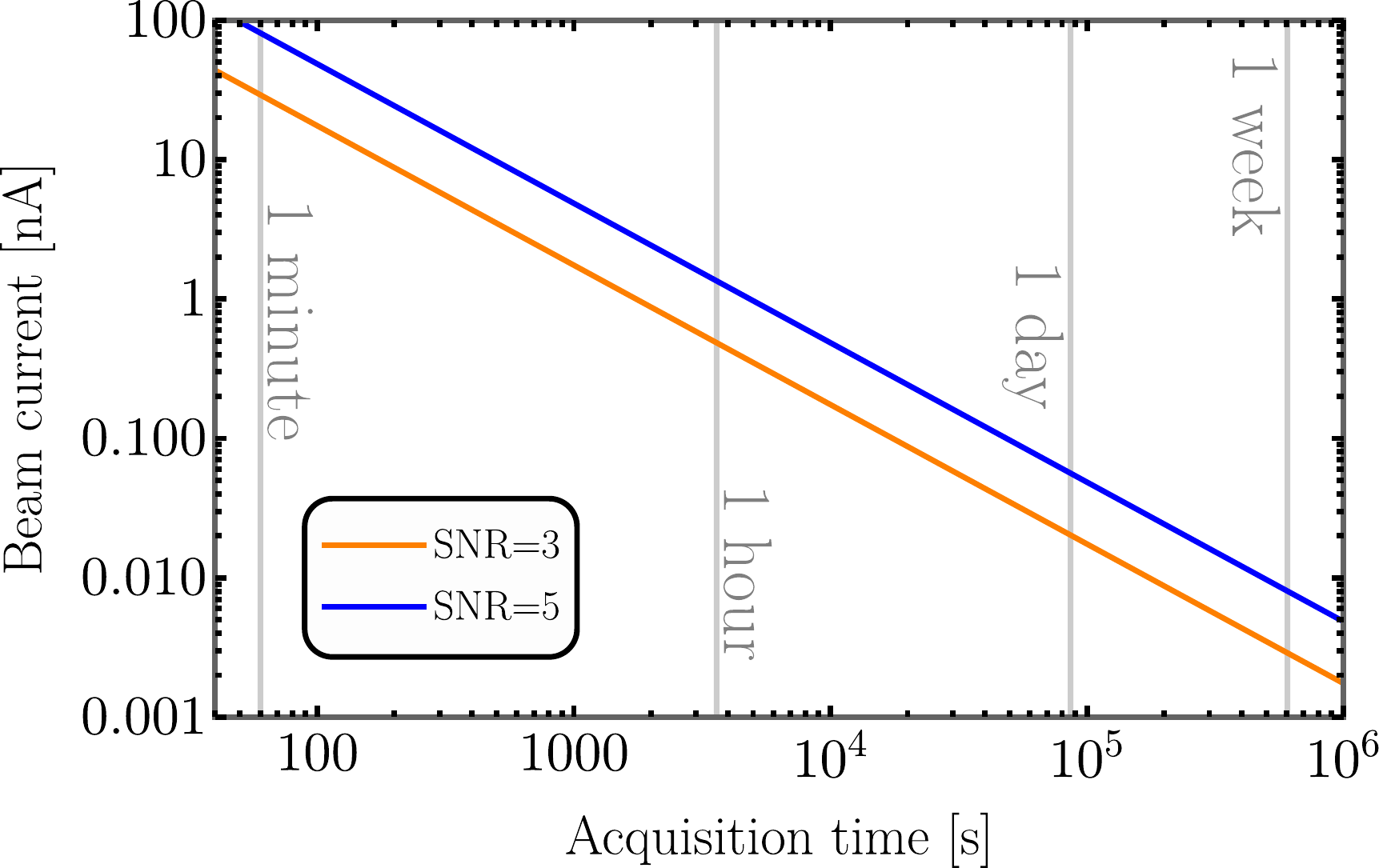}
	\caption{
 Log-log plots of the electron beam current $i_b$ as a function of the acquisition time $t_a$---computed from Eq.~(\ref{Eq:SNR})---producing signal-to-noise ratios (SNR) of 3 and 5, for bcc Fe of 16.072~nm thickness at \review{1100}~K.
 }
	\label{Fig:currentvstimeSNR}
\end{figure}

A typical criterion for successful detection conditions in \review{(S)TEM} is to have at least $\text{SNR}=3$~\cite{carter2016transmission}, while in general signal processing $\text{SNR}=5$ criterion is used~\cite{Rose1973}. Therefore, using Eq.~(\ref{Eq:SNR}) for the sample of 16.072~nm thickness at \review{1100}~K, we present in Fig.~\ref{Fig:currentvstimeSNR} log-log plots of $i_b$ as a function of $t_a$ giving $\text{SNR}=3$ and $\text{SNR}=5$. 
%
%
%
%
%

It is worth mentioning that existing \review{(S)TEM} machines vary from some of which have low $i_b$, on the order of pA, to others capable of routinely working well above 1~nA \cite{Houdellier2015107,shao2018high,konings2020}. Therefore, it can be appreciated in Fig.~\ref{Fig:currentvstimeSNR} that our calculations predict large yet \review{attainable} acquisition times for realistic conditions (below one hour for beam currents over 1~nA). For example, it is predicted that a $\text{SNR}=3$ could be obtained in less than 30~min. using a 1~nA electron probe.
%
%
Also, it is worth noticing that for tens of pA, the acquisition times are less than one day. \review{Clearly, experiments will encounter constraints related to sample drift and the maximum allowable electron dose before beam-induced damage appears~\cite{egerton2011electron}. Nevertheless, 
employing a scanning probe across an area of uniform sample thickness and orientation may help 
to alleviate these limitations. Another potential path to increase the relative strength of $I_\text{MDS}$ could be an exploration of the use of patterned apertures~\cite{zeltmann}. 
Furthermore, in specific systems, magnon and phonon signals may be sufficiently separated in energy, allowing individual filtering or mapping (including as a function of temperature) by incorporating energy resolution through EELS, offering another level to disentangle their contributions.}


    %
    %
\subsection{\label{sec:eels}Discussion}

We have shown that, in our simulations, there are optimal detection conditions at which $I_\text{MDS}$ can be statistically significant at \review{conceivable} acquisition times. Therefore, there might exist experimental configurations allowing for temperature-aided detection of MDS in \review{(S)TEM}.

However, we have not put forward an explicit method to separate the two contributions, $I_\text{TDS}$ and $I_\text{MDS}$. A possible starting point in this endeavor could rely on the difference in the concavity of the signals at the optimal conditions, pointed out in the previous subsection (i.e., the behavior around \review{1100}~K for the 16.072~nm curves in the top panel of Fig.~\ref{Fig:SNRandTDS} and in the bottom panel of Fig.~\ref{Fig:adf}). In particular, it could be useful to fit with Gaussian functions the experimental ADF-collected signals at different temperatures to detect the $I_\text{MDS}$ peaks. 
%
%
Moreover, rotating the samples' magnetization could be attempted, since it may influence $I_\text{MDS}$, leaving $I_\text{TDS}$ unaltered.

Nevertheless, independently of any specific detection strategy, we consider that the main finding of our work is that, under \review{accomplishable} measurement conditions, the $I_\text{MDS}$ signal can become a statistically significant contribution to the total scattering intensity within a suitably chosen detector.

To further bridge the gap toward successful MDS detection in STEM, it would be valuable to perform energy-resolved STEM studies. For that matter, a theoretical methodology allowing for MDS simulations with energy resolution would be of utmost relevance. Nevertheless, the findings reported in this work could likely help to establish optimal conditions for STEM-EELS MDS studies, both theoretical and experimental.



%
%
\section{\label{sec:conclusions}Conclusions}

We have presented a study of simulated magnon diffuse scattering (MDS) in bcc Fe samples, of different thicknesses and temperatures, to explore the possibility of temperature-aided MDS detection in scanning transmission electron microscopy (STEM). An annular dark-field (ADF) detector that collects the relevant MDS signal surrounding the central Bragg disk [illustrated by the green dashed lines in Fig.~\ref{Fig:nodwfdiff}(a)] has been employed.

It was found that when the effects of the atomic vibrations are neglected, the MDS signal $I_\text{MDS}$ grows approximately linearly up to the Curie temperature ($T_\text{C}$) of Fe, presenting a much less-varying behavior for higher temperatures. Also, instead of increasing monotonically as a function of thickness, the MDS signal displayed an alternating behavior due to dynamical diffraction.

When \review{the effects of the atomic vibrations are incorporated through the complex atomic electrostatic potential}, the linear growth of $I_\text{MDS}$ gives place to a \review{different non-linear} behavior, which presents a predominant peak \review{for the sample of thickness 16.072~nm at 1100~K}. 
%
%

In contrast, the thermal diffuse scattering (TDS), due to the atomic vibrations (phonons), presented a signal ($I_\text{TDS}$) that grows more linearly\review{---in all the temperature range considered---}than \review{the corresponding $I_\text{MDS}$ [including the complex $V(\mathbf{r})$]}, but still displayed appreciable dynamical diffraction effects. Moreover, it was found that $I_\text{TDS}$ was five orders of magnitude greater than the corresponding $I_\text{MDS}$. Nevertheless, an analysis of the signal-to-noise ratio (SNR) showed that under \review{realizable} measurement conditions, the $I_\text{MDS}$ signal can become a statistically significant contribution to the total scattering intensity. In particular, we found that $\text{SNR} \geq 3$ \review{could} be achieved with existing \review{(S)TEM} machines in less than 30~min. of data acquisition for a bcc Fe sample of 16.072~nm thickness at \review{1100}~K.


%
%
\begin{acknowledgments}
We acknowledge the support of the Swedish Research Council, Olle Engkvist’s foundation, Carl Trygger's Foundation, Knut and Alice Wallenberg Foundation, and eSSENCE for financial support. 
The simulations were enabled by resources provided by the National Academic Infrastructure for Supercomputing in Sweden (NAISS) and the Swedish National Infrastructure for Computing (SNIC) at NSC Centre partially funded by the Swedish Research Council through grant agreements no. 2022-06725 and no. 2018-05973.
SuperSTEM is the National Research Facility for Advanced Electron Microscopy funded by the Engineering and Physical Sciences Research Council (EPSRC). We acknowledge financial support from the Engineering and Physical Sciences Research Council (EPSRC) via Grant No. EP/V048767/1 and Royal Society Grant No. IES/R1/211016. J.C.I. acknowledges the support of the Center for Nanophase Materials Sciences, which is a DOE Office of Science User Facility. \review{We would like to especially thank the anonymous referees who have offered invaluable and meticulous feedback on an earlier iteration of this manuscript. In particular, we are grateful for the suggestion to include the effects of an absorptive potential.}
\end{acknowledgments}

%
%

\appendix 

%
%

\section{\label{app:statistical} Convergence in terms of the averaging of snapshots}

The accuracy of the inelastic signals $I_\text{ine}(T)$ computed with Eq.~(\ref{Eq:Iinelastic}) depends, in particular, on the number of snapshots $N_s$ considered for the averaging process. A higher $N_s$ produces a more converged value of $I_\text{ine}(T)$.

Given a fixed $N_s$, it is relevant to estimate the degree of accuracy of the computed signals. A rough estimation can be achieved by exploiting the fact that the elastic signal $I_\text{ela}(T)$ should consist of only Bragg disks. 

%
\begin{figure}[h]
	\centering
	\includegraphics[width=\linewidth]{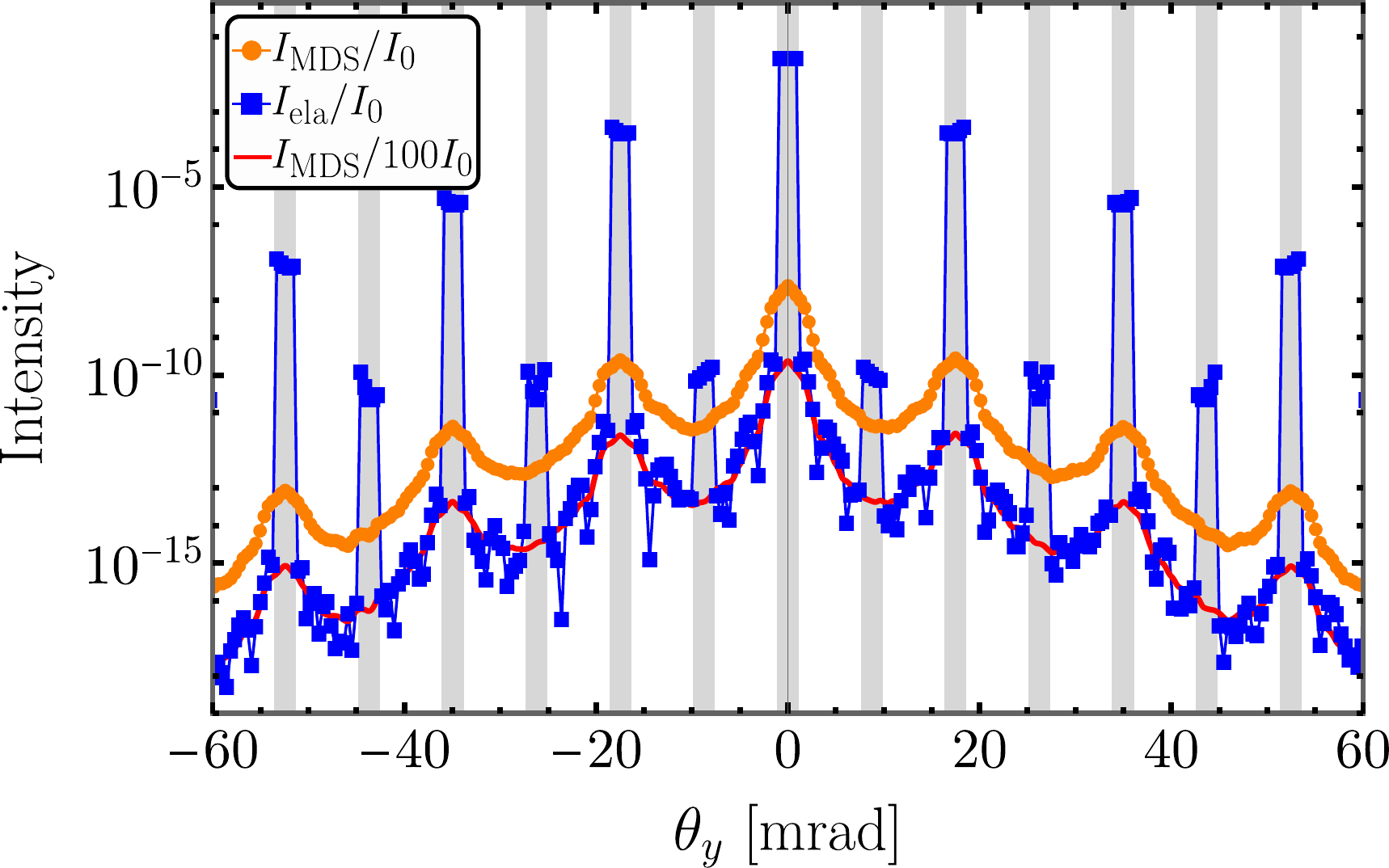}
	\caption{Elastic and magnon diffuse scattering signals---$I_\text{ela}$ and $I_\text{MDS}$ (divided by the total intensity $I_0$ of the incident beam integrated over the whole diffraction plane), respectively [see Eq.~(\ref{Eq:Iinelastic})]---for bcc Fe of
 \review{16.072~nm thickness at 1100~K, including the effects of atomic vibrations through the complex $V(\mathbf{r})$}
 %
 and using the cutoff distance $r_\text{cut}=1$~nm. A curve of $I_\text{MDS}/I_0$ divided by 100 is included for comparison with $I_\text{ela}$. The position and width of the Bragg disks are indicated by the vertical gray bars.}
	\label{Fig:statisticalMDS}
\end{figure}
%

Take for example Fig.~\ref{Fig:statisticalMDS}, showing signals for bcc Fe of 
\review{16.072~nm thickness at 1100~K---including the complex $V(\mathbf{r})$ to account for the effects of atomic vibrations, and using the cutoff distance $r_\text{cut}=1$~nm (for information about $V(\mathbf{r})$ and $r_\text{cut}$, please refer to the last two paragraphs of Section~\ref{sec:methods}).}
Specifically, in Fig.~\ref{Fig:statisticalMDS} we show $I_\text{ela}/I_0$ and $I_\text{MDS}/I_0$ (where $I_0$ denotes the intensity of the incident electron beam integrated over the whole diffraction plane) along a vertical profile through the center of the diffraction plane, i.e., as a function of the scattering angle $\theta_y$, with $\theta_x=0$. We have included vertical gray bars indicating the position and width of the Bragg disks.

It can be appreciated in Fig.~\ref{Fig:statisticalMDS} that there is a non-zero elastic signal between the Bragg disks. This signal, called hereafter $I_\text{error}^{N_s}$, goes to zero as $N_s$ increases. Hence, from Eq.~(\ref{Eq:Iinelastic}), computing $I_\text{MDS}$ with a given $N_s$ gives an error on the order of $I_\text{error}^{N_s}$ in the resulting MDS signal (outside the Bragg disks). In this work, in which $N_s=101$, and in particular in Fig.~\ref{Fig:statisticalMDS}, $I_\text{error}^{N_s=101}$ is about two orders of magnitude smaller than the corresponding $I_\text{MDS}$. We include $I_\text{MDS}/(100I_0)$ in Fig.~\ref{Fig:statisticalMDS} (red curve) to help illustrate this result.


\review{
%
%

\section{\label{app:meansquaredisplacements} Complex atomic electrostatic potential}

We have employed the parameterization for the complex atomic electrostatic potential $V(\mathbf{r})$ developed by Peng et al.~\cite{DUDAREV199586,Peng1996}, which includes absorption effects due to TDS. Specifically, we utilized Eq.~(8) of Ref.~\cite{DUDAREV199586}, corresponding to the following formula for the complex electrostatic potential energy $U(\mathbf{r}) = e V(\mathbf{r})$ (where $e$ represents the elemental charge):
%
%
\begin{equation}
U(\mathbf{r})
=
-
\frac{2\pi\hbar^2}{m_0}
\sum_a
\sum_{n=1}^5
U_n\left(\mathbf{r} - \mathbf{R}_a\right)
.
\label{Eq:PengU}
\end{equation}
%
%
In Eq.~(\ref{Eq:PengU}), $\mathbf{R}_a$ denotes the equilibrium position of atom $a$, $\sum_a$ denotes the sum over the atomic positions in the crystal, $m_0$ is the electron rest mass, and $\hbar$ is Planck's reduced constant. The complex function $U_n(\mathbf{r})$ can be expressed in terms of its real (elastic) and imaginary (absorptive) parts as
%
%
\begin{equation}
U_n(\mathbf{r})
=
U_n^\text{(Re)}\left(\mathbf{r}\right)
+
i
U_n^\text{(abs)}\left(\mathbf{r}\right)
,
\label{Eq:PengUn}
\end{equation}
%
%
with $i=\sqrt{-1}$, and
%
%
\begin{equation}
U_n^\text{(Re)}
\!
\left(\mathbf{r}\right)
=
a_n^\text{(Re)}
\!
\left[
\frac{4\pi}{b_n^\text{(Re)}\! + \!B}
\right]^\frac{3}{2}
\!\!\!
\exp
\!
\left[
\frac{-4 \pi^2 r^2}{b_n^\text{(Re)}\! + \!B}
\right]
,
\label{Eq:PengunRe}
\end{equation}
%
%
%
%
\begin{equation}
U_n^\text{(abs)}
\!
\left(\mathbf{r}\right)
\!
=
\!
a_n^\text{(abs)}
\!\!
\left[
\frac{4\pi}{b_n^\text{(abs)}\! + \!\frac{B}{2}}
\right]^\frac{3}{2}
\!\!\!\!
\exp
\!\!
\left[
\frac{-4 \pi^2 r^2}{b_n^\text{(abs)}\! + \!\frac{B}{2}}
\right]
.
\label{Eq:Penguntds}
\end{equation}
%
%

 The constants $a_n^\text{(Re)}$, $b_n^\text{(Re)}$, $a_n^\text{(abs)}$, and $b_n^\text{(abs)}$ are real-valued fitting parameters~\footnote{In Ref.~\cite{DUDAREV199586}, the parameters $a_n^\text{(abs)}$ and $b_n^\text{(abs)}$ are called $a_n^\text{(TDS)}$ and $b_n^\text{(TDS)}$, respectively.}. The symbol $B$ represents the isotropic $B$-factor of the Debye-Waller factor for bcc Fe, $B_\text{iso} = 8\pi^2\langle \mathbf{u}^2\rangle_T/3$~\cite{zeiger2023lessons,Malica:ae5064}, in which $\langle \mathbf{u}^2 \rangle_T$ is the mean-squared displacement at temperature $T$. The values of $\langle \mathbf{u}^2 \rangle_T$ used for the determination of $B$ factors at temperature $T$ are given in Table~\ref{tab:meansquaredisplacements}, alongside the lattice parameter of bcc Fe, as obtained by molecular dynamics calculations. 

%
%
%
\begin{table}[h]
\caption{\label{tab:meansquaredisplacements}%
Mean-squared displacement $\langle \mathbf{u}^2 \rangle_T$---used for the implementation of Debye-Waller factors---and bcc Fe lattice parameter $a$, at temperature $T$, used for TDS calculations.
}
\begin{ruledtabular}
\begin{tabular}{ccccccc}
$T$ [K] & $a$ [\AA] & $\langle \mathbf{u}^2 \rangle_T$ [\AA$^2$] & & $T$ [K] & $a$ [\AA] & $\langle \mathbf{u}^2 \rangle_T$ [\AA$^2$]
\vspace{0.5mm}
\\
\colrule
\vspace{-3mm} &  &  & &  &  &  \\
 100 & 2.855389 & 0.004693 & & 1000 & 2.881882 & 0.061313 \\
 200 & 2.856854 & 0.009678 & & 1100 & 2.886190 & 0.069974 \\
 300 & 2.858989 & 0.014849 & & 1200 & 2.890794 & 0.079865 \\
 400 & 2.861494 & 0.020589 & & 1300 & 2.895650 & 0.090492 \\
 500 & 2.864269 & 0.026535 & & 1400 & 2.900800 & 0.102204 \\
 600 & 2.867294 & 0.032765 & & 1500 & 2.906238 & 0.115404 \\
 700 & 2.870572 & 0.038693 & & 1600 & 2.912003 & 0.130029 \\
 800 & 2.874080 & 0.045774 & & 1700 & 2.918126 & 0.148881 \\
 900 & 2.877857 & 0.053119 & & & & \\
\end{tabular}
\end{ruledtabular}
\end{table}
%

In our study, we employed the numerical value $2\pi\hbar^2/m_0 = 47.87798$~\AA$^2$eV~\cite{DUDAREV199586} in Eq.~(\ref{Eq:PengU}). Furthermore, for the parameters in Eqs.~(\ref{Eq:PengunRe}) and (\ref{Eq:Penguntds}) we used the values of Tables~\ref{tab:aPengreal}, \ref{tab:bPengreal}, and \ref{tab:Pengim}.
%
%
\begin{table}[h]
\caption{\label{tab:aPengreal}%
Values of the parameters $a_n^\text{(Re)}$ of Eq.~(\ref{Eq:PengunRe}) for bcc Fe. Taken from Table~3 of Ref.~\cite{Peng1996}.
}
\begin{ruledtabular}
\begin{tabular}{ccccc}
$a_1^\text{(Re)}$ [\AA] & $a_2^\text{(Re)}$ [\AA] & $a_3^\text{(Re)}$ [\AA] & $a_4^\text{(Re)}$ [\AA] & $a_5^\text{(Re)}$ [\AA]
\vspace{0.5mm}
\\
\colrule
\vspace{-3mm} &  &  &   &    \\
 0.1929 & 0.8239 & 1.8689 &  2.3694 & 1.9060 \\
\end{tabular}
\end{ruledtabular}
\end{table}
%
%
%
%
%
%
\begin{table}[h]
\caption{\label{tab:bPengreal}%
Values of the parameters $b_n^\text{(Re)}$ of Eq.~(\ref{Eq:PengunRe}) for bcc Fe. Taken from Table~3 of Ref.~\cite{Peng1996}.
}
\begin{ruledtabular}
\begin{tabular}{ccccc}
$b_1^\text{(Re)}$ [\AA$^2$] & $b_2^\text{(Re)}$ [\AA$^2$] & $b_3^\text{(Re)}$ [\AA$^2$] & $b_4^\text{(Re)}$ [\AA$^2$] & $b_5^\text{(Re)}$ [\AA$^2$]
\vspace{0.5mm}
\\
\colrule
\vspace{-3mm} &  &  &   &    \\
 0.1087 & 1.0806 & 4.7637 &  22.8500 & 76.7309 \\
\end{tabular}
\end{ruledtabular}
\end{table}
%
%
%
%
%

%
\begin{table*}
\caption{\label{tab:Pengim}%
Values of the parameters $a_n^\text{(abs)}$ (in [\AA] units) and $b_n^\text{(abs)}$ (in [\AA$^2$] units) of Eq.~(\ref{Eq:Penguntds}) for bcc Fe determined by the fitting procedure described in Refs.~\cite{DUDAREV199586,Peng1996} for a 200~kV electron probe. The DWF $B$-factor was computed from the $\langle \mathbf{u}^2 \rangle_T$ values of Table~\ref{tab:meansquaredisplacements}.  The rightmost column shows the standard deviation $\sigma$ of our fit, determined using Eq.~(11) of Ref.~\cite{Peng1996}.
}
\begin{ruledtabular}
\begin{tabular}{ccccccccccccc}
$T$ [K] 
& 
$a_1^\text{(abs)}$ & $a_2^\text{(abs)}$ & $a_3^\text{(abs)}$ & $a_4^\text{(abs)}$ & $a_5^\text{(abs)}$  
&
$b_1^\text{(abs)}$ & $b_2^\text{(abs)}$ & $b_3^\text{(abs)}$ & $b_4^\text{(abs)}$ & $b_5^\text{(abs)}$
&
$B$ [\AA$^2$] 
&
$\sigma$ [\AA] 
\vspace{0.5mm}
\\
\hline & \\[-2mm]
100	& 0.01733	&	0.00383	&	0.09533	&	-0.05651	&	-0.01291	&	1.07730	&	4.85519	&	0.18916	&	0.22153	&	0.05072	&	0.123515	&	0.858$ \times 10^{-5}$  \\
200	& 0.00750	&	0.03280	&	10.92648	&	-10.88299	&	-0.00728	&	5.02555	&	1.16514	&	0.16692	&	0.16659	&	0.04500	&	0.254715	&	1.689$ \times 10^{-5}$  \\
300	& 0.01106	&	0.04708	&	0.06690	&	-0.02151	&	-0.00392	&	5.16181	&	1.24349	&	0.28783	&	0.12304	&	0.03903	&	0.390810	&	2.471$ \times 10^{-5}$  \\
400	& 0.01409	&	0.05949	&	0.06826	&	-0.01895	&	-0.00243	&	5.44356	&	1.36831	&	0.40155	&	0.12765	&	0.03306	&	0.541881	&	3.196$ \times 10^{-5}$  \\
500	& 0.01576	&	0.06723	&	0.07671	&	-0.01878	&	-0.00230	&	5.91156	&	1.55839	&	0.53568	&	0.14657	&	0.03487	&	0.698373	&	3.699$ \times 10^{-5}$  \\
600	& 0.01517	&	0.06845	&	0.09338	&	-0.00233	&	-0.01962	&	6.77315	&	1.87670	&	0.69916	&	0.03835	&	0.17037	&	0.862340	&	3.869$ \times 10^{-5}$  \\
700	& 0.06572	&	0.01236	&	0.11428	&	-0.00241	&	-0.02114	&	2.32300	&	8.24889	&	0.86116	&	0.04238	&	0.19603	&	1.018359	&	3.621$ \times 10^{-5}$  \\
800	& 0.00902	&	0.06413	&	0.13665	&	-0.00261	&	-0.02376	&	10.75286	&	2.87720	&	1.03051	&	0.04849	&	0.23155	&	1.204723	&	2.914$ \times 10^{-5}$  \\
900	& 0.06472	&	0.00728	&	0.15457	&	-0.00279	&	-0.02691	&	3.32041	&	13.27529	&	1.17903	&	0.05471	&	0.26989	&	1.398036	&	2.089$ \times 10^{-5}$  \\
1000 & 0.06607	&	0.00645	&	0.17089	&	-0.00290	&	-0.03024	&	3.70311	&	15.55992	&	1.33059	&	0.06062	&	0.30992	&	1.613693	&	1.420$ \times 10^{-5}$  \\
1100 & 0.00607	&	0.06724	&	0.18583	&	-0.00296	&	-0.03326	&	17.47419	&	4.04853	&	1.48567	&	0.06600	&	0.34819	&	1.841642	&	1.294$ \times 10^{-5}$  \\
1200 & 0.00588	&	0.06775	&	0.20123	&	-0.00299	&	-0.03603	&	19.25031	&	4.41662	&	1.66213	&	0.07158	&	0.38765	&	2.101963	&	1.755$ \times 10^{-5}$  \\
1300 & 0.06716	&	0.00577	&	0.21670	&	-0.00301	&	-0.03834	&	4.81115	&	20.86226	&	1.85335	&	0.07734	&	0.42646	&	2.381654	&	2.440$ \times 10^{-5}$  \\
1400 & 0.06522	&	0.00568	&	0.23302	&	-0.00306	&	-0.04030	&	5.26642	&	22.43683	&	2.06637	&	0.08369	&	0.46646	&	2.689901	&	3.189$ \times 10^{-5}$  \\
1500 & 0.06181	&	0.00556	&	0.25073	&	-0.00313	&	-0.04203	&	5.82167	&	24.08175	&	2.30793	&	0.09110	&	0.50960	&	3.037312	&	3.965$ \times 10^{-5}$  \\
1600 & 0.05720	&	0.00537	&	0.26944	&	-0.00325	&	-0.04359	&	6.50055	&	25.84895	&	2.57465	&	0.09981	&	0.55638	&	3.422226	&	4.733$ \times 10^{-5}$  \\
1700 & 0.30563	&	0.04107	&	-0.04288	&	-0.00036	&	-0.00470	&	2.99216	&	10.03400	&	0.63216	&	0.02431	&	0.17428	&	3.918391	&	7.502$ \times 10^{-5}$   \\
\end{tabular}
\end{ruledtabular}
\end{table*}
%
%
%
%

The values of the parameters in Tables~\ref{tab:aPengreal} and \ref{tab:bPengreal} are reproduced from Table~3 of Ref.~\cite{Peng1996}. To obtain the values of Tables~\ref{tab:Pengim}, we implemented the fitting methodology described in Refs.~\cite{DUDAREV199586,Peng1996} in an in-house program, employing (i) the values of Tables~\ref{tab:aPengreal} and \ref{tab:bPengreal}, (ii) the values of $B$ computed from $\langle \mathbf{u}^2 \rangle_T$ in Table~\ref{tab:meansquaredisplacements}, and (iii) a 200~kV acceleration voltage. We have included in Table~\ref{tab:Pengim} the standard deviation $\sigma$ of our fit, computed as Eq.~(11) of Ref.~\cite{Peng1996}.
%
%
%
%
%
%
\begin{figure}[h]
	\centering
	\includegraphics[width=\linewidth]{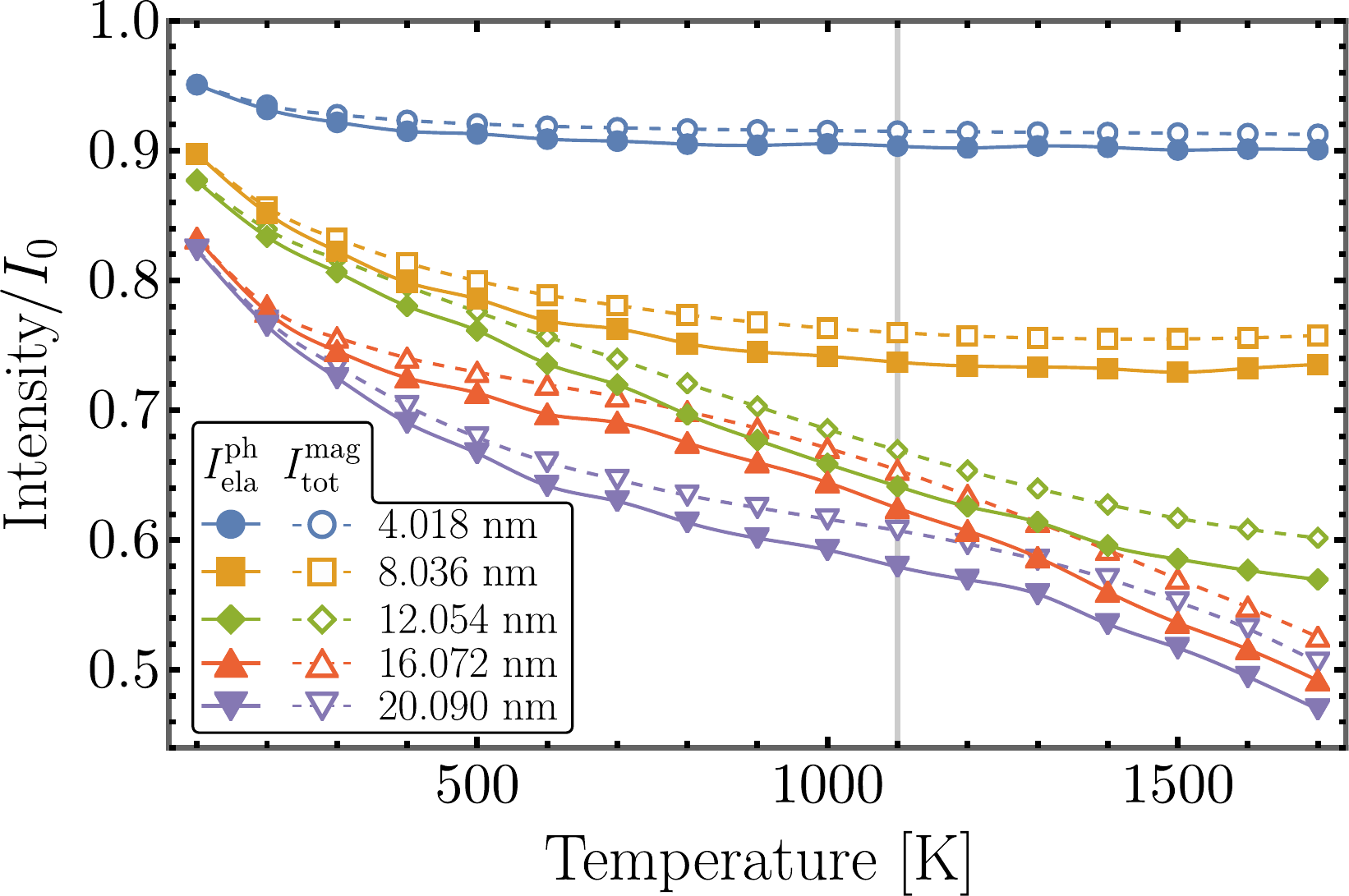}
	\caption{\review{Comparison of the elastic signal $I_\text{ela}^\text{ph}$ obtained in FPMS simulations with the total signal $I_\text{tot}^\text{mag}$ from FMMS simulations---including the effects of atomic vibrations through the complex potential $V(\mathbf{r})$---as a function of temperature for various specimen thicknesses. Both signals are integrated over the entire diffraction plane and normalized by the total incident beam intensity $I_0$, which is also integrated over the entire diffraction plane.}}
	\label{Fig:integralcomparison}
\end{figure}
%

To demonstrate the effectiveness of the complex potential in replicating TDS absorption, we display in Fig.~\ref{Fig:integralcomparison} the elastic signal obtained in the FPMS simulations, denoted as $I_\text{ela}^\text{ph}$, integrated over the entire diffraction plane. This signal is shown as a function of temperature for various specimen thicknesses, along with the integrated total signal from FMMS simulations, denoted as $I_\text{tot}^\text{mag}$. Both signals are normalized by the total incident beam intensity, denoted as $I_0$, integrated over the entire diffraction plane. The percentage difference between both signals, given by $100\times\lvert I_\text{ela}^\text{ph} - I_\text{tot}^\text{mag}\rvert / \lvert I_\text{ela}^\text{ph} \rvert$, remains below $8\%$ for all cases considered. Consequently, the reduction of the integrated intensity in the elastic channel observed in FPMS simulations, attributed to phonons, closely matches the corresponding total signal decrease in FMMS simulations due to the absorptive component of the complex potential. 

Notably, in Fig.~\ref{Fig:integralcomparison}, it can be appreciated that $I_\text{ela}^\text{ph} \leq I_\text{tot}^\text{mag}$, indicating that the depletion of $I_\text{ela}^\text{ph}$ consistently exceeds that of $I_\text{tot}^\text{mag}$. This behavior likely stems from the contrasting approaches used to address atomic motion in the two methodologies.  The complex potential model assumes uncorrelated atomic motion, whereas the FPMS methodology fully considers these correlations. Consequently, there could be non-local (off-diagonal) contributions to the absorptive potential resulting from correlated atomic motion, which are neglected in a local absorptive potential (see Fig.~3 in Ref.~\cite{AllenPhysRevB}). These contributions could explain the slightly increased depletion of the elastic channel in FPMS as compared to FMMS (see Fig.~3 and the related discussion in Ref.~\cite{AllenMandM}).



%

}

%
%

\section{\label{app:cutoff} Convergence in terms of \review{the cutoff distance}}

The microscopic electromagnetic fields $V(\mathbf{r})$, $\mathbf{B}(\mathbf{r})$, and $\mathbf{A}(\mathbf{r})$ of an atom vanish as the distance from the atom increases. Therefore, it is customary to define a cutoff distance $r_\text{cut}$ above which these fields are set to zero to economize computational resources in crystals simulations. This establishes a compromise between the precision of a calculation and its computational resources demand.

%
\begin{figure}[h]
	\centering
	\includegraphics[width=\linewidth]{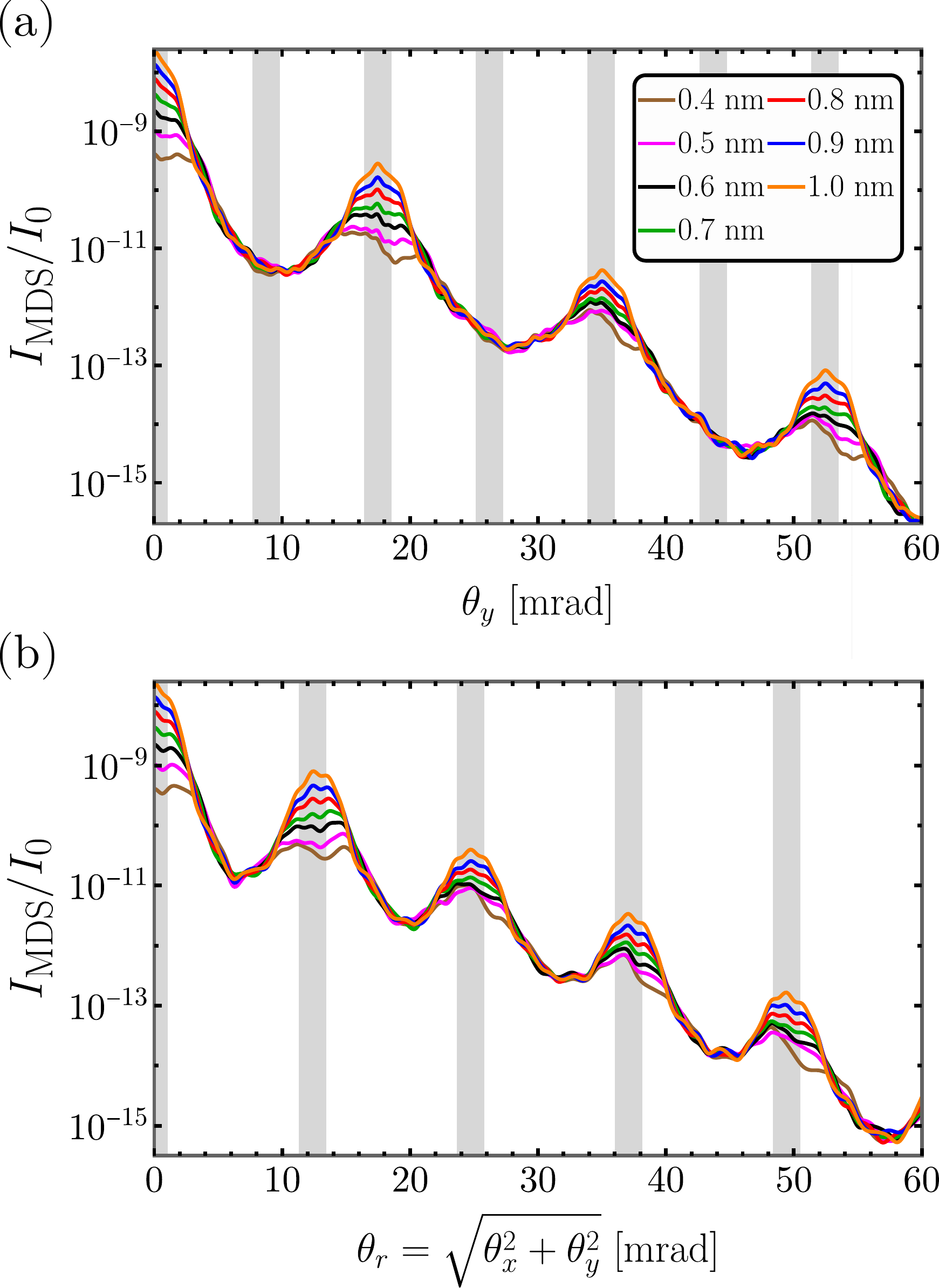}
	\caption{Convergence of the magnon diffuse scattering signal $I_\text{MDS}$ (divided by the intensity $I_0$ of the incident beam \review{integrated over the whole diffraction plane}) for different cutoff distances $r_\text{cut}$---indicated in the inset of the top plot [Fig.~\ref{Fig:cutoff}(a)]---for bcc Fe of \review{16.072}~nm thickness at \review{1100}~K, including the Debye-Waller factor (DWF). (a) Vertical [i.e., as a function of the scattering angle $\theta_y$ for fixed $\theta_x=0$] and (b) diagonal [i.e., as a function of $\theta_r = \sqrt{\theta_x^2 + \theta_y^2}$, with $\theta_x = \theta_y$] profiles through the origin of the diffraction plane. The gray bars indicate the position and width of the Bragg disks. 
}
	\label{Fig:cutoff}
\end{figure}
%

To show the convergence level of our calculations, in Fig.~\ref{Fig:cutoff} we show MDS profiles (in logarithmic scale) for different values of $r_\text{cut}$, as a function of the scattering angle $\theta_r$, for bcc Fe of \review{16.072~nm thickness, at 1100~K, fully including the complex $V(\mathbf{r})$}. Specifically, in Fig.~\ref{Fig:cutoff}(a) we show the vertical profile of $I_\text{MDS}/I_0$ ($I_0$ denotes the intensity of the incident electron beam) through the origin of the diffraction plane---that is, as a function of $\theta_y$ for fixed $\theta_x=0$. Meanwhile, Fig.~\ref{Fig:cutoff}(b) shows a diagonal profile of $I_\text{MDS}/I_0$ through the origin of the diffraction plane---i.e., as a function of $\theta_r = \sqrt{\theta_x^2 + \theta_y^2}$ with $\theta_x = \theta_y$. We have included vertical gray bars indicating the position and width of the Bragg disks in both Figs.~\ref{Fig:cutoff}(a) and ~\ref{Fig:cutoff}(b).

It can be observed in these figures that the MDS signal decreases as the scattering angle increases, presenting maxima near the position of the high-intensity Bragg disks (in agreement with Figs.~\ref{Fig:nodwfdiff}(a), ~\ref{Fig:nodwfdiff}(b), and~\ref{Fig:statisticalMDS}). Moreover, the highest MDS signal is located around the central Bragg spot (at zero scattering angle), in conformity with Ref.~\cite{mendis2022}. Therefore, in our study, we considered an annular dark-field (ADF) detector~\cite{kirkland2010} of outer collection semiangle 7~mrad and inner collection semiangle 2~mrad [represented by the green dashed lines in Fig.~\ref{Fig:nodwfdiff}(a)] to avoid all Bragg disks.

In particular, well converged MDS signals within the considered ADF detector can be obtained with $r_\text{cut} = 0.4$~nm, as shown Figs.~\ref{Fig:cutoff}(a) and ~\ref{Fig:cutoff}(b). However, to better resolve the MDS near and within the Bragg disks, a higher $r_\text{cut}$ is necessary. 



%
%


%
\bibliographystyle{apsrev4-2}

\bibliography{references}

\providecommand{\noopsort}[1]{}\providecommand{\singleletter}[1]{#1}%
\begin{thebibliography}{53}%
\makeatletter
\providecommand \@ifxundefined [1]{%
 \@ifx{#1\undefined}
}%
\providecommand \@ifnum [1]{%
 \ifnum #1\expandafter \@firstoftwo
 \else \expandafter \@secondoftwo
 \fi
}%
\providecommand \@ifx [1]{%
 \ifx #1\expandafter \@firstoftwo
 \else \expandafter \@secondoftwo
 \fi
}%
\providecommand \natexlab [1]{#1}%
\providecommand \enquote  [1]{``#1''}%
\providecommand \bibnamefont  [1]{#1}%
\providecommand \bibfnamefont [1]{#1}%
\providecommand \citenamefont [1]{#1}%
\providecommand \href@noop [0]{\@secondoftwo}%
\providecommand \href [0]{\begingroup \@sanitize@url \@href}%
\providecommand \@href[1]{\@@startlink{#1}\@@href}%
\providecommand \@@href[1]{\endgroup#1\@@endlink}%
\providecommand \@sanitize@url [0]{\catcode `\\12\catcode `\$12\catcode
  `\&12\catcode `\#12\catcode `\^12\catcode `\_12\catcode `\%12\relax}%
\providecommand \@@startlink[1]{}%
\providecommand \@@endlink[0]{}%
\providecommand \url  [0]{\begingroup\@sanitize@url \@url }%
\providecommand \@url [1]{\endgroup\@href {#1}{\urlprefix }}%
\providecommand \urlprefix  [0]{URL }%
\providecommand \Eprint [0]{\href }%
\providecommand \doibase [0]{https://doi.org/}%
\providecommand \selectlanguage [0]{\@gobble}%
\providecommand \bibinfo  [0]{\@secondoftwo}%
\providecommand \bibfield  [0]{\@secondoftwo}%
\providecommand \translation [1]{[#1]}%
\providecommand \BibitemOpen [0]{}%
\providecommand \bibitemStop [0]{}%
\providecommand \bibitemNoStop [0]{.\EOS\space}%
\providecommand \EOS [0]{\spacefactor3000\relax}%
\providecommand \BibitemShut  [1]{\csname bibitem#1\endcsname}%
\let\auto@bib@innerbib\@empty
\bibitem [{\citenamefont {Carter}\ and\ \citenamefont
  {Williams}(2009)}]{carter2009transmission}%
  \BibitemOpen
  \bibfield  {author} {\bibinfo {author} {\bibfnamefont {C.~B.}\ \bibnamefont
  {Carter}}\ and\ \bibinfo {author} {\bibfnamefont {D.~B.}\ \bibnamefont
  {Williams}},\ }\href
  {https://doi.org/https://doi.org/10.1007/978-0-387-76501-3} {\emph {\bibinfo
  {title} {Transmission Electron Microscopy: A Textbook for Materials
  Science}}}\ (\bibinfo  {publisher} {Springer},\ \bibinfo {year}
  {2009})\BibitemShut {NoStop}%
\bibitem [{\citenamefont {Krivanek}\ \emph {et~al.}(2014)\citenamefont
  {Krivanek}, \citenamefont {Lovejoy}, \citenamefont {Dellby}, \citenamefont
  {Aoki}, \citenamefont {Carpenter}, \citenamefont {Rez}, \citenamefont
  {Soignard}, \citenamefont {Zhu}, \citenamefont {Batson}, \citenamefont
  {Lagos}, \citenamefont {Egerton},\ and\ \citenamefont
  {Crozier}}]{krivanek2014nature}%
  \BibitemOpen
  \bibfield  {author} {\bibinfo {author} {\bibfnamefont {O.~L.}\ \bibnamefont
  {Krivanek}}, \bibinfo {author} {\bibfnamefont {T.~C.}\ \bibnamefont
  {Lovejoy}}, \bibinfo {author} {\bibfnamefont {N.}~\bibnamefont {Dellby}},
  \bibinfo {author} {\bibfnamefont {T.}~\bibnamefont {Aoki}}, \bibinfo {author}
  {\bibfnamefont {R.}~\bibnamefont {Carpenter}}, \bibinfo {author}
  {\bibfnamefont {P.}~\bibnamefont {Rez}}, \bibinfo {author} {\bibfnamefont
  {E.}~\bibnamefont {Soignard}}, \bibinfo {author} {\bibfnamefont
  {J.}~\bibnamefont {Zhu}}, \bibinfo {author} {\bibfnamefont {P.~E.}\
  \bibnamefont {Batson}}, \bibinfo {author} {\bibfnamefont {M.~J.}\
  \bibnamefont {Lagos}}, \bibinfo {author} {\bibfnamefont {R.~F.}\ \bibnamefont
  {Egerton}},\ and\ \bibinfo {author} {\bibfnamefont {P.~A.}\ \bibnamefont
  {Crozier}},\ }\href {https://doi.org/https://doi.org/10.1038/nature13870}
  {\bibfield  {journal} {\bibinfo  {journal} {Nature}\ }\textbf {\bibinfo
  {volume} {514}},\ \bibinfo {pages} {209} (\bibinfo {year}
  {2014})}\BibitemShut {NoStop}%
\bibitem [{\citenamefont {Krivanek}\ \emph {et~al.}(2019)\citenamefont
  {Krivanek}, \citenamefont {Dellby}, \citenamefont {Hachtel}, \citenamefont
  {Idrobo}, \citenamefont {Hotz}, \citenamefont {Plotkin-Swing}, \citenamefont
  {Bacon}, \citenamefont {Bleloch}, \citenamefont {Corbin}, \citenamefont
  {Hoffman}, \citenamefont {Meyer},\ and\ \citenamefont
  {Lovejoy}}]{krivanek2019ultramicroscopy}%
  \BibitemOpen
  \bibfield  {author} {\bibinfo {author} {\bibfnamefont {O.}~\bibnamefont
  {Krivanek}}, \bibinfo {author} {\bibfnamefont {N.}~\bibnamefont {Dellby}},
  \bibinfo {author} {\bibfnamefont {J.}~\bibnamefont {Hachtel}}, \bibinfo
  {author} {\bibfnamefont {J.-C.}\ \bibnamefont {Idrobo}}, \bibinfo {author}
  {\bibfnamefont {M.}~\bibnamefont {Hotz}}, \bibinfo {author} {\bibfnamefont
  {B.}~\bibnamefont {Plotkin-Swing}}, \bibinfo {author} {\bibfnamefont
  {N.}~\bibnamefont {Bacon}}, \bibinfo {author} {\bibfnamefont
  {A.}~\bibnamefont {Bleloch}}, \bibinfo {author} {\bibfnamefont
  {G.}~\bibnamefont {Corbin}}, \bibinfo {author} {\bibfnamefont
  {M.}~\bibnamefont {Hoffman}}, \bibinfo {author} {\bibfnamefont
  {C.}~\bibnamefont {Meyer}},\ and\ \bibinfo {author} {\bibfnamefont
  {T.}~\bibnamefont {Lovejoy}},\ }\href
  {https://doi.org/https://doi.org/10.1016/j.ultramic.2018.12.006} {\bibfield
  {journal} {\bibinfo  {journal} {Ultramicroscopy}\ }\textbf {\bibinfo {volume}
  {203}},\ \bibinfo {pages} {60} (\bibinfo {year} {2019})}\BibitemShut
  {NoStop}%
\bibitem [{\citenamefont {Dellby}\ \emph {et~al.}(2020)\citenamefont {Dellby},
  \citenamefont {Lovejoy}, \citenamefont {Corbin}, \citenamefont {Johnson},
  \citenamefont {Hayner}, \citenamefont {Hoffman}, \citenamefont {Hrncrik},
  \citenamefont {Plotkin-Swing}, \citenamefont {Taylor},\ and\ \citenamefont
  {Krivanek}}]{dellby2020ultra}%
  \BibitemOpen
  \bibfield  {author} {\bibinfo {author} {\bibfnamefont {N.}~\bibnamefont
  {Dellby}}, \bibinfo {author} {\bibfnamefont {T.}~\bibnamefont {Lovejoy}},
  \bibinfo {author} {\bibfnamefont {G.}~\bibnamefont {Corbin}}, \bibinfo
  {author} {\bibfnamefont {N.}~\bibnamefont {Johnson}}, \bibinfo {author}
  {\bibfnamefont {R.}~\bibnamefont {Hayner}}, \bibinfo {author} {\bibfnamefont
  {M.}~\bibnamefont {Hoffman}}, \bibinfo {author} {\bibfnamefont
  {P.}~\bibnamefont {Hrncrik}}, \bibinfo {author} {\bibfnamefont
  {B.}~\bibnamefont {Plotkin-Swing}}, \bibinfo {author} {\bibfnamefont
  {D.}~\bibnamefont {Taylor}},\ and\ \bibinfo {author} {\bibfnamefont
  {O.}~\bibnamefont {Krivanek}},\ }\href
  {https://doi.org/10.1017/S1431927620019406} {\bibfield  {journal} {\bibinfo
  {journal} {Microsc. Microanal.}\ }\textbf {\bibinfo {volume} {26}},\ \bibinfo
  {pages} {1804} (\bibinfo {year} {2020})}\BibitemShut {NoStop}%
\bibitem [{\citenamefont {Dellby}\ \emph {et~al.}(2022)\citenamefont {Dellby},
  \citenamefont {Krivanek}, \citenamefont {Bacon}, \citenamefont {Corbin},
  \citenamefont {Johnson}, \citenamefont {Hayner}, \citenamefont {Hrncrik},
  \citenamefont {Plotkin-Swing}, \citenamefont {Taylor}, \citenamefont
  {Szilaygi},\ and\ \citenamefont {Lovejoy}}]{dellby2022multi}%
  \BibitemOpen
  \bibfield  {author} {\bibinfo {author} {\bibfnamefont {N.}~\bibnamefont
  {Dellby}}, \bibinfo {author} {\bibfnamefont {O.}~\bibnamefont {Krivanek}},
  \bibinfo {author} {\bibfnamefont {N.}~\bibnamefont {Bacon}}, \bibinfo
  {author} {\bibfnamefont {G.}~\bibnamefont {Corbin}}, \bibinfo {author}
  {\bibfnamefont {N.}~\bibnamefont {Johnson}}, \bibinfo {author} {\bibfnamefont
  {R.}~\bibnamefont {Hayner}}, \bibinfo {author} {\bibfnamefont
  {P.}~\bibnamefont {Hrncrik}}, \bibinfo {author} {\bibfnamefont
  {B.}~\bibnamefont {Plotkin-Swing}}, \bibinfo {author} {\bibfnamefont
  {D.}~\bibnamefont {Taylor}}, \bibinfo {author} {\bibfnamefont
  {Z.}~\bibnamefont {Szilaygi}},\ and\ \bibinfo {author} {\bibfnamefont
  {T.}~\bibnamefont {Lovejoy}},\ }\href
  {https://doi.org/10.1017/S1431927622010017} {\bibfield  {journal} {\bibinfo
  {journal} {Microsc. Microanal.}\ }\textbf {\bibinfo {volume} {28}},\ \bibinfo
  {pages} {2640} (\bibinfo {year} {2022})}\BibitemShut {NoStop}%
\bibitem [{\citenamefont {Lagos}\ \emph {et~al.}(2022)\citenamefont {Lagos},
  \citenamefont {Bicket}, \citenamefont {Mousavi~M.},\ and\ \citenamefont
  {Botton}}]{lagos2022advances}%
  \BibitemOpen
  \bibfield  {author} {\bibinfo {author} {\bibfnamefont {M.~J.}\ \bibnamefont
  {Lagos}}, \bibinfo {author} {\bibfnamefont {I.~C.}\ \bibnamefont {Bicket}},
  \bibinfo {author} {\bibfnamefont {S.~S.}\ \bibnamefont {Mousavi~M.}},\ and\
  \bibinfo {author} {\bibfnamefont {G.~A.}\ \bibnamefont {Botton}},\ }\href
  {https://doi.org/10.1093/jmicro/dfab050} {\bibfield  {journal} {\bibinfo
  {journal} {Microscopy}\ }\textbf {\bibinfo {volume} {71}},\ \bibinfo {pages}
  {i174} (\bibinfo {year} {2022})}\BibitemShut {NoStop}%
\bibitem [{\citenamefont {Lyon}\ \emph {et~al.}(2021)\citenamefont {Lyon},
  \citenamefont {Bergman}, \citenamefont {Zeiger}, \citenamefont
  {Kepaptsoglou}, \citenamefont {Ramasse}, \citenamefont {Idrobo},\ and\
  \citenamefont {Rusz}}]{Lyon2021PRB}%
  \BibitemOpen
  \bibfield  {author} {\bibinfo {author} {\bibfnamefont {K.}~\bibnamefont
  {Lyon}}, \bibinfo {author} {\bibfnamefont {A.}~\bibnamefont {Bergman}},
  \bibinfo {author} {\bibfnamefont {P.}~\bibnamefont {Zeiger}}, \bibinfo
  {author} {\bibfnamefont {D.}~\bibnamefont {Kepaptsoglou}}, \bibinfo {author}
  {\bibfnamefont {Q.~M.}\ \bibnamefont {Ramasse}}, \bibinfo {author}
  {\bibfnamefont {J.~C.}\ \bibnamefont {Idrobo}},\ and\ \bibinfo {author}
  {\bibfnamefont {J.}~\bibnamefont {Rusz}},\ }\href
  {https://doi.org/10.1103/PhysRevB.104.214418} {\bibfield  {journal} {\bibinfo
   {journal} {Phys. Rev. B}\ }\textbf {\bibinfo {volume} {104}},\ \bibinfo
  {pages} {214418} (\bibinfo {year} {2021})}\BibitemShut {NoStop}%
\bibitem [{\citenamefont {Mendis}(2021)}]{mendis2021}%
  \BibitemOpen
  \bibfield  {author} {\bibinfo {author} {\bibfnamefont {B.}~\bibnamefont
  {Mendis}},\ }\href
  {https://doi.org/https://doi.org/10.1016/j.ultramic.2021.113390} {\bibfield
  {journal} {\bibinfo  {journal} {Ultramicroscopy}\ }\textbf {\bibinfo {volume}
  {230}},\ \bibinfo {pages} {113390} (\bibinfo {year} {2021})}\BibitemShut
  {NoStop}%
\bibitem [{\citenamefont {Mendis}(2022)}]{mendis2022}%
  \BibitemOpen
  \bibfield  {author} {\bibinfo {author} {\bibfnamefont {B.}~\bibnamefont
  {Mendis}},\ }\href
  {https://doi.org/https://doi.org/10.1016/j.ultramic.2022.113548} {\bibfield
  {journal} {\bibinfo  {journal} {Ultramicroscopy}\ }\textbf {\bibinfo {volume}
  {239}},\ \bibinfo {pages} {113548} (\bibinfo {year} {2022})}\BibitemShut
  {NoStop}%
\bibitem [{\citenamefont {Eriksson}\ \emph {et~al.}(2017)\citenamefont
  {Eriksson}, \citenamefont {Bergman}, \citenamefont {Bergqvist},\ and\
  \citenamefont {Hellsvik}}]{Eriksson2017}%
  \BibitemOpen
  \bibfield  {author} {\bibinfo {author} {\bibfnamefont {O.}~\bibnamefont
  {Eriksson}}, \bibinfo {author} {\bibfnamefont {A.}~\bibnamefont {Bergman}},
  \bibinfo {author} {\bibfnamefont {L.}~\bibnamefont {Bergqvist}},\ and\
  \bibinfo {author} {\bibfnamefont {J.}~\bibnamefont {Hellsvik}},\ }\href
  {https://doi.org/10.1093/oso/9780198788669.001.0001} {\emph {\bibinfo {title}
  {Atomistic Spin Dynamics: Foundations and Applications}}}\ (\bibinfo
  {publisher} {Oxford University Press},\ \bibinfo {year} {2017})\BibitemShut
  {NoStop}%
\bibitem [{\citenamefont {Kittel}\ and\ \citenamefont
  {McEuen}(2018)}]{kittel2018introduction}%
  \BibitemOpen
  \bibfield  {author} {\bibinfo {author} {\bibfnamefont {C.}~\bibnamefont
  {Kittel}}\ and\ \bibinfo {author} {\bibfnamefont {P.}~\bibnamefont
  {McEuen}},\ }\href@noop {} {\emph {\bibinfo {title} {Introduction to Solid
  State Physics}}}\ (\bibinfo  {publisher} {John Wiley \& Sons},\ \bibinfo
  {year} {2018})\BibitemShut {NoStop}%
\bibitem [{\citenamefont {Mohn}(2006)}]{mohn2006magnetism}%
  \BibitemOpen
  \bibfield  {author} {\bibinfo {author} {\bibfnamefont {P.}~\bibnamefont
  {Mohn}},\ }\href {https://doi.org/https://doi.org/10.1007/3-540-30981-0}
  {\emph {\bibinfo {title} {Magnetism in the Solid State}}}\ (\bibinfo
  {publisher} {Springer Science \& Business Media},\ \bibinfo {year}
  {2006})\BibitemShut {NoStop}%
\bibitem [{\citenamefont {\ifmmode \check{Z}\else
  \v{Z}\fi{}uti\ifmmode~\acute{c}\else \'{c}\fi{}}\ \emph
  {et~al.}(2004)\citenamefont {\ifmmode \check{Z}\else
  \v{Z}\fi{}uti\ifmmode~\acute{c}\else \'{c}\fi{}}, \citenamefont {Fabian},\
  and\ \citenamefont {Das~Sarma}}]{zutic2004}%
  \BibitemOpen
  \bibfield  {author} {\bibinfo {author} {\bibfnamefont {I.}~\bibnamefont
  {\ifmmode \check{Z}\else \v{Z}\fi{}uti\ifmmode~\acute{c}\else \'{c}\fi{}}},
  \bibinfo {author} {\bibfnamefont {J.}~\bibnamefont {Fabian}},\ and\ \bibinfo
  {author} {\bibfnamefont {S.}~\bibnamefont {Das~Sarma}},\ }\href
  {https://doi.org/10.1103/RevModPhys.76.323} {\bibfield  {journal} {\bibinfo
  {journal} {Rev. Mod. Phys.}\ }\textbf {\bibinfo {volume} {76}},\ \bibinfo
  {pages} {323} (\bibinfo {year} {2004})}\BibitemShut {NoStop}%
\bibitem [{\citenamefont {Pulizzi}(2012)}]{pulizzi2012spintronics}%
  \BibitemOpen
  \bibfield  {author} {\bibinfo {author} {\bibfnamefont {F.}~\bibnamefont
  {Pulizzi}},\ }\href {https://doi.org/10.1038/nmat3327} {\bibfield  {journal}
  {\bibinfo  {journal} {Nat. Mater.}\ }\textbf {\bibinfo {volume} {11}},\
  \bibinfo {pages} {367} (\bibinfo {year} {2012})}\BibitemShut {NoStop}%
\bibitem [{\citenamefont {Barman}\ \emph {et~al.}(2021)\citenamefont {Barman},
  \citenamefont {Gubbiotti}, \citenamefont {Ladak}, \citenamefont {Adeyeye},
  \citenamefont {Krawczyk}, \citenamefont {Gräfe}, \citenamefont {Adelmann},
  \citenamefont {Cotofana}, \citenamefont {Naeemi}, \citenamefont {Vasyuchka},
  \citenamefont {Hillebrands}, \citenamefont {Nikitov}, \citenamefont {Yu},
  \citenamefont {Grundler}, \citenamefont {Sadovnikov}, \citenamefont
  {Grachev}, \citenamefont {Sheshukova}, \citenamefont {Duquesne},
  \citenamefont {Marangolo}, \citenamefont {Csaba}, \citenamefont {Porod},
  \citenamefont {Demidov}, \citenamefont {Urazhdin}, \citenamefont
  {Demokritov}, \citenamefont {Albisetti}, \citenamefont {Petti}, \citenamefont
  {Bertacco}, \citenamefont {Schultheiss}, \citenamefont {Kruglyak},
  \citenamefont {Poimanov}, \citenamefont {Sahoo}, \citenamefont {Sinha},
  \citenamefont {Yang}, \citenamefont {Münzenberg}, \citenamefont {Moriyama},
  \citenamefont {Mizukami}, \citenamefont {Landeros}, \citenamefont {Gallardo},
  \citenamefont {Carlotti}, \citenamefont {Kim}, \citenamefont {Stamps},
  \citenamefont {Camley}, \citenamefont {Rana}, \citenamefont {Otani},
  \citenamefont {Yu}, \citenamefont {Yu}, \citenamefont {Bauer}, \citenamefont
  {Back}, \citenamefont {Uhrig}, \citenamefont {Dobrovolskiy}, \citenamefont
  {Budinska}, \citenamefont {Qin}, \citenamefont {van Dijken}, \citenamefont
  {Chumak}, \citenamefont {Khitun}, \citenamefont {Nikonov}, \citenamefont
  {Young}, \citenamefont {Zingsem},\ and\ \citenamefont
  {Winklhofer}}]{Barman2021}%
  \BibitemOpen
  \bibfield  {author} {\bibinfo {author} {\bibfnamefont {A.}~\bibnamefont
  {Barman}}, \bibinfo {author} {\bibfnamefont {G.}~\bibnamefont {Gubbiotti}},
  \bibinfo {author} {\bibfnamefont {S.}~\bibnamefont {Ladak}}, \bibinfo
  {author} {\bibfnamefont {A.~O.}\ \bibnamefont {Adeyeye}}, \bibinfo {author}
  {\bibfnamefont {M.}~\bibnamefont {Krawczyk}}, \bibinfo {author}
  {\bibfnamefont {J.}~\bibnamefont {Gräfe}}, \bibinfo {author} {\bibfnamefont
  {C.}~\bibnamefont {Adelmann}}, \bibinfo {author} {\bibfnamefont
  {S.}~\bibnamefont {Cotofana}}, \bibinfo {author} {\bibfnamefont
  {A.}~\bibnamefont {Naeemi}}, \bibinfo {author} {\bibfnamefont {V.~I.}\
  \bibnamefont {Vasyuchka}}, \bibinfo {author} {\bibfnamefont {B.}~\bibnamefont
  {Hillebrands}}, \bibinfo {author} {\bibfnamefont {S.~A.}\ \bibnamefont
  {Nikitov}}, \bibinfo {author} {\bibfnamefont {H.}~\bibnamefont {Yu}},
  \bibinfo {author} {\bibfnamefont {D.}~\bibnamefont {Grundler}}, \bibinfo
  {author} {\bibfnamefont {A.~V.}\ \bibnamefont {Sadovnikov}}, \bibinfo
  {author} {\bibfnamefont {A.~A.}\ \bibnamefont {Grachev}}, \bibinfo {author}
  {\bibfnamefont {S.~E.}\ \bibnamefont {Sheshukova}}, \bibinfo {author}
  {\bibfnamefont {J.-Y.}\ \bibnamefont {Duquesne}}, \bibinfo {author}
  {\bibfnamefont {M.}~\bibnamefont {Marangolo}}, \bibinfo {author}
  {\bibfnamefont {G.}~\bibnamefont {Csaba}}, \bibinfo {author} {\bibfnamefont
  {W.}~\bibnamefont {Porod}}, \bibinfo {author} {\bibfnamefont {V.~E.}\
  \bibnamefont {Demidov}}, \bibinfo {author} {\bibfnamefont {S.}~\bibnamefont
  {Urazhdin}}, \bibinfo {author} {\bibfnamefont {S.~O.}\ \bibnamefont
  {Demokritov}}, \bibinfo {author} {\bibfnamefont {E.}~\bibnamefont
  {Albisetti}}, \bibinfo {author} {\bibfnamefont {D.}~\bibnamefont {Petti}},
  \bibinfo {author} {\bibfnamefont {R.}~\bibnamefont {Bertacco}}, \bibinfo
  {author} {\bibfnamefont {H.}~\bibnamefont {Schultheiss}}, \bibinfo {author}
  {\bibfnamefont {V.~V.}\ \bibnamefont {Kruglyak}}, \bibinfo {author}
  {\bibfnamefont {V.~D.}\ \bibnamefont {Poimanov}}, \bibinfo {author}
  {\bibfnamefont {S.}~\bibnamefont {Sahoo}}, \bibinfo {author} {\bibfnamefont
  {J.}~\bibnamefont {Sinha}}, \bibinfo {author} {\bibfnamefont
  {H.}~\bibnamefont {Yang}}, \bibinfo {author} {\bibfnamefont {M.}~\bibnamefont
  {Münzenberg}}, \bibinfo {author} {\bibfnamefont {T.}~\bibnamefont
  {Moriyama}}, \bibinfo {author} {\bibfnamefont {S.}~\bibnamefont {Mizukami}},
  \bibinfo {author} {\bibfnamefont {P.}~\bibnamefont {Landeros}}, \bibinfo
  {author} {\bibfnamefont {R.~A.}\ \bibnamefont {Gallardo}}, \bibinfo {author}
  {\bibfnamefont {G.}~\bibnamefont {Carlotti}}, \bibinfo {author}
  {\bibfnamefont {J.-V.}\ \bibnamefont {Kim}}, \bibinfo {author} {\bibfnamefont
  {R.~L.}\ \bibnamefont {Stamps}}, \bibinfo {author} {\bibfnamefont {R.~E.}\
  \bibnamefont {Camley}}, \bibinfo {author} {\bibfnamefont {B.}~\bibnamefont
  {Rana}}, \bibinfo {author} {\bibfnamefont {Y.}~\bibnamefont {Otani}},
  \bibinfo {author} {\bibfnamefont {W.}~\bibnamefont {Yu}}, \bibinfo {author}
  {\bibfnamefont {T.}~\bibnamefont {Yu}}, \bibinfo {author} {\bibfnamefont
  {G.~E.~W.}\ \bibnamefont {Bauer}}, \bibinfo {author} {\bibfnamefont
  {C.}~\bibnamefont {Back}}, \bibinfo {author} {\bibfnamefont {G.~S.}\
  \bibnamefont {Uhrig}}, \bibinfo {author} {\bibfnamefont {O.~V.}\ \bibnamefont
  {Dobrovolskiy}}, \bibinfo {author} {\bibfnamefont {B.}~\bibnamefont
  {Budinska}}, \bibinfo {author} {\bibfnamefont {H.}~\bibnamefont {Qin}},
  \bibinfo {author} {\bibfnamefont {S.}~\bibnamefont {van Dijken}}, \bibinfo
  {author} {\bibfnamefont {A.~V.}\ \bibnamefont {Chumak}}, \bibinfo {author}
  {\bibfnamefont {A.}~\bibnamefont {Khitun}}, \bibinfo {author} {\bibfnamefont
  {D.~E.}\ \bibnamefont {Nikonov}}, \bibinfo {author} {\bibfnamefont {I.~A.}\
  \bibnamefont {Young}}, \bibinfo {author} {\bibfnamefont {B.~W.}\ \bibnamefont
  {Zingsem}},\ and\ \bibinfo {author} {\bibfnamefont {M.}~\bibnamefont
  {Winklhofer}},\ }\href {https://doi.org/10.1088/1361-648X/abec1a} {\bibfield
  {journal} {\bibinfo  {journal} {J. Phys. Condens. Matter}\ }\textbf {\bibinfo
  {volume} {33}},\ \bibinfo {pages} {413001} (\bibinfo {year}
  {2021})}\BibitemShut {NoStop}%
\bibitem [{\citenamefont {Vollmer}\ \emph {et~al.}(2004)\citenamefont
  {Vollmer}, \citenamefont {Etzkorn}, \citenamefont {{Anil Kumar}},
  \citenamefont {Ibach},\ and\ \citenamefont {Kirschner}}]{VOLLMER20042126}%
  \BibitemOpen
  \bibfield  {author} {\bibinfo {author} {\bibfnamefont {R.}~\bibnamefont
  {Vollmer}}, \bibinfo {author} {\bibfnamefont {M.}~\bibnamefont {Etzkorn}},
  \bibinfo {author} {\bibfnamefont {P.}~\bibnamefont {{Anil Kumar}}}, \bibinfo
  {author} {\bibfnamefont {H.}~\bibnamefont {Ibach}},\ and\ \bibinfo {author}
  {\bibfnamefont {J.}~\bibnamefont {Kirschner}},\ }\href
  {https://doi.org/https://doi.org/10.1016/j.jmmm.2003.12.506} {\bibfield
  {journal} {\bibinfo  {journal} {J. Magn. Magn. Mater.}\ }\textbf {\bibinfo
  {volume} {272-276}},\ \bibinfo {pages} {2126} (\bibinfo {year}
  {2004})}\BibitemShut {NoStop}%
\bibitem [{\citenamefont {Zakeri}\ \emph {et~al.}(2013)\citenamefont {Zakeri},
  \citenamefont {Zhang},\ and\ \citenamefont {Kirschner}}]{Zakeri2013}%
  \BibitemOpen
  \bibfield  {author} {\bibinfo {author} {\bibfnamefont {K.}~\bibnamefont
  {Zakeri}}, \bibinfo {author} {\bibfnamefont {Y.}~\bibnamefont {Zhang}},\ and\
  \bibinfo {author} {\bibfnamefont {J.}~\bibnamefont {Kirschner}},\ }\href
  {https://doi.org/https://doi.org/10.1016/j.elspec.2012.06.009} {\bibfield
  {journal} {\bibinfo  {journal} {J. Electron. Spectrosc. Relat. Phenom.}\
  }\textbf {\bibinfo {volume} {189}},\ \bibinfo {pages} {157} (\bibinfo {year}
  {2013})}\BibitemShut {NoStop}%
\bibitem [{\citenamefont {Ibach}\ \emph {et~al.}(2017)\citenamefont {Ibach},
  \citenamefont {Bocquet}, \citenamefont {Sforzini}, \citenamefont {Soubatch},\
  and\ \citenamefont {Tautz}}]{ibach2017electron}%
  \BibitemOpen
  \bibfield  {author} {\bibinfo {author} {\bibfnamefont {H.}~\bibnamefont
  {Ibach}}, \bibinfo {author} {\bibfnamefont {F.~C.}\ \bibnamefont {Bocquet}},
  \bibinfo {author} {\bibfnamefont {J.}~\bibnamefont {Sforzini}}, \bibinfo
  {author} {\bibfnamefont {S.}~\bibnamefont {Soubatch}},\ and\ \bibinfo
  {author} {\bibfnamefont {F.~S.}\ \bibnamefont {Tautz}},\ }\href
  {https://doi.org/https://doi.org/10.1063/1.4977529} {\bibfield  {journal}
  {\bibinfo  {journal} {Rev. Sci. Instrum.}\ }\textbf {\bibinfo {volume}
  {88}},\ \bibinfo {pages} {033903} (\bibinfo {year} {2017})}\BibitemShut
  {NoStop}%
\bibitem [{\citenamefont {Idrobo}\ \emph {et~al.}(2018)\citenamefont {Idrobo},
  \citenamefont {Lupini}, \citenamefont {Feng}, \citenamefont {Unocic},
  \citenamefont {Walden}, \citenamefont {Gardiner}, \citenamefont {Lovejoy},
  \citenamefont {Dellby}, \citenamefont {Pantelides},\ and\ \citenamefont
  {Krivanek}}]{IdroboPhysRevLett.120.095901}%
  \BibitemOpen
  \bibfield  {author} {\bibinfo {author} {\bibfnamefont {J.~C.}\ \bibnamefont
  {Idrobo}}, \bibinfo {author} {\bibfnamefont {A.~R.}\ \bibnamefont {Lupini}},
  \bibinfo {author} {\bibfnamefont {T.}~\bibnamefont {Feng}}, \bibinfo {author}
  {\bibfnamefont {R.~R.}\ \bibnamefont {Unocic}}, \bibinfo {author}
  {\bibfnamefont {F.~S.}\ \bibnamefont {Walden}}, \bibinfo {author}
  {\bibfnamefont {D.~S.}\ \bibnamefont {Gardiner}}, \bibinfo {author}
  {\bibfnamefont {T.~C.}\ \bibnamefont {Lovejoy}}, \bibinfo {author}
  {\bibfnamefont {N.}~\bibnamefont {Dellby}}, \bibinfo {author} {\bibfnamefont
  {S.~T.}\ \bibnamefont {Pantelides}},\ and\ \bibinfo {author} {\bibfnamefont
  {O.~L.}\ \bibnamefont {Krivanek}},\ }\href
  {https://doi.org/10.1103/PhysRevLett.120.095901} {\bibfield  {journal}
  {\bibinfo  {journal} {Phys. Rev. Lett.}\ }\textbf {\bibinfo {volume} {120}},\
  \bibinfo {pages} {095901} (\bibinfo {year} {2018})}\BibitemShut {NoStop}%
\bibitem [{\citenamefont {Lagos}\ and\ \citenamefont
  {Batson}(2018)}]{lagos2018thermometry}%
  \BibitemOpen
  \bibfield  {author} {\bibinfo {author} {\bibfnamefont {M.~J.}\ \bibnamefont
  {Lagos}}\ and\ \bibinfo {author} {\bibfnamefont {P.~E.}\ \bibnamefont
  {Batson}},\ }\href {https://doi.org/10.1021/acs.nanolett.8b01791} {\bibfield
  {journal} {\bibinfo  {journal} {Nano Lett.}\ }\textbf {\bibinfo {volume}
  {18}},\ \bibinfo {pages} {4556} (\bibinfo {year} {2018})}\BibitemShut
  {NoStop}%
\bibitem [{\citenamefont {Kikkawa}\ and\ \citenamefont
  {Kimoto}(2022)}]{KikkawaPhysRevB.106.195431}%
  \BibitemOpen
  \bibfield  {author} {\bibinfo {author} {\bibfnamefont {J.}~\bibnamefont
  {Kikkawa}}\ and\ \bibinfo {author} {\bibfnamefont {K.}~\bibnamefont
  {Kimoto}},\ }\href {https://doi.org/10.1103/PhysRevB.106.195431} {\bibfield
  {journal} {\bibinfo  {journal} {Phys. Rev. B}\ }\textbf {\bibinfo {volume}
  {106}},\ \bibinfo {pages} {195431} (\bibinfo {year} {2022})}\BibitemShut
  {NoStop}%
\bibitem [{\citenamefont {Wehmeyer}\ \emph {et~al.}(2018)\citenamefont
  {Wehmeyer}, \citenamefont {Bustillo}, \citenamefont {Minor},\ and\
  \citenamefont {Dames}}]{wehmeyer}%
  \BibitemOpen
  \bibfield  {author} {\bibinfo {author} {\bibfnamefont {G.}~\bibnamefont
  {Wehmeyer}}, \bibinfo {author} {\bibfnamefont {K.~C.}\ \bibnamefont
  {Bustillo}}, \bibinfo {author} {\bibfnamefont {A.~M.}\ \bibnamefont
  {Minor}},\ and\ \bibinfo {author} {\bibfnamefont {C.}~\bibnamefont {Dames}},\
  }\href {https://doi.org/10.1063/1.5066111} {\bibfield  {journal} {\bibinfo
  {journal} {Appl. Phys. Lett.}\ }\textbf {\bibinfo {volume} {113}},\ \bibinfo
  {pages} {253101} (\bibinfo {year} {2018})}\BibitemShut {NoStop}%
\bibitem [{\citenamefont {Loane}\ \emph {et~al.}(1991)\citenamefont {Loane},
  \citenamefont {Xu},\ and\ \citenamefont {Silcox}}]{Loane1991}%
  \BibitemOpen
  \bibfield  {author} {\bibinfo {author} {\bibfnamefont {R.~F.}\ \bibnamefont
  {Loane}}, \bibinfo {author} {\bibfnamefont {P.}~\bibnamefont {Xu}},\ and\
  \bibinfo {author} {\bibfnamefont {J.}~\bibnamefont {Silcox}},\ }\href
  {https://doi.org/10.1107/S0108767391000375} {\bibfield  {journal} {\bibinfo
  {journal} {Acta Cryst. A}\ }\textbf {\bibinfo {volume} {47}},\ \bibinfo
  {pages} {267} (\bibinfo {year} {1991})}\BibitemShut {NoStop}%
\bibitem [{\citenamefont {Kirkland}(2010)}]{kirkland2010}%
  \BibitemOpen
  \bibfield  {author} {\bibinfo {author} {\bibfnamefont {E.~J.}\ \bibnamefont
  {Kirkland}},\ }\href {https://doi.org/10.1007/978-1-4419-6533-2} {\emph
  {\bibinfo {title} {Advanced Computing in Electron Microscopy}}}\ (\bibinfo
  {publisher} {Springer},\ \bibinfo {year} {2010})\BibitemShut {NoStop}%
\bibitem [{\citenamefont {Edstr\"om}\ \emph
  {et~al.}(2016{\natexlab{a}})\citenamefont {Edstr\"om}, \citenamefont {Lubk},\
  and\ \citenamefont {Rusz}}]{Edstrom2016PRB}%
  \BibitemOpen
  \bibfield  {author} {\bibinfo {author} {\bibfnamefont {A.}~\bibnamefont
  {Edstr\"om}}, \bibinfo {author} {\bibfnamefont {A.}~\bibnamefont {Lubk}},\
  and\ \bibinfo {author} {\bibfnamefont {J.}~\bibnamefont {Rusz}},\ }\href
  {https://doi.org/10.1103/PhysRevB.94.174414} {\bibfield  {journal} {\bibinfo
  {journal} {Phys. Rev. B}\ }\textbf {\bibinfo {volume} {94}},\ \bibinfo
  {pages} {174414} (\bibinfo {year} {2016}{\natexlab{a}})}\BibitemShut
  {NoStop}%
\bibitem [{\citenamefont {Edstr\"om}\ \emph
  {et~al.}(2016{\natexlab{b}})\citenamefont {Edstr\"om}, \citenamefont {Lubk},\
  and\ \citenamefont {Rusz}}]{Edstrom2016PRL}%
  \BibitemOpen
  \bibfield  {author} {\bibinfo {author} {\bibfnamefont {A.}~\bibnamefont
  {Edstr\"om}}, \bibinfo {author} {\bibfnamefont {A.}~\bibnamefont {Lubk}},\
  and\ \bibinfo {author} {\bibfnamefont {J.}~\bibnamefont {Rusz}},\ }\href
  {https://doi.org/10.1103/PhysRevLett.116.127203} {\bibfield  {journal}
  {\bibinfo  {journal} {Phys. Rev. Lett.}\ }\textbf {\bibinfo {volume} {116}},\
  \bibinfo {pages} {127203} (\bibinfo {year} {2016}{\natexlab{b}})}\BibitemShut
  {NoStop}%
\bibitem [{\citenamefont {Zeiger}\ and\ \citenamefont
  {Rusz}(2020)}]{Zeiger2020}%
  \BibitemOpen
  \bibfield  {author} {\bibinfo {author} {\bibfnamefont {P.~M.}\ \bibnamefont
  {Zeiger}}\ and\ \bibinfo {author} {\bibfnamefont {J.}~\bibnamefont {Rusz}},\
  }\href {https://doi.org/10.1103/PhysRevLett.124.025501} {\bibfield  {journal}
  {\bibinfo  {journal} {Phys. Rev. Lett.}\ }\textbf {\bibinfo {volume} {124}},\
  \bibinfo {pages} {025501} (\bibinfo {year} {2020})}\BibitemShut {NoStop}%
\bibitem [{\citenamefont {{Van Dyck}}(2009)}]{VanDyck2009}%
  \BibitemOpen
  \bibfield  {author} {\bibinfo {author} {\bibfnamefont {D.}~\bibnamefont {{Van
  Dyck}}},\ }\href
  {https://doi.org/https://doi.org/10.1016/j.ultramic.2009.01.001} {\bibfield
  {journal} {\bibinfo  {journal} {Ultramicroscopy}\ }\textbf {\bibinfo {volume}
  {109}},\ \bibinfo {pages} {677} (\bibinfo {year} {2009})}\BibitemShut
  {NoStop}%
\bibitem [{\citenamefont {Forbes}\ \emph {et~al.}(2010)\citenamefont {Forbes},
  \citenamefont {Martin}, \citenamefont {Findlay}, \citenamefont {D'Alfonso},\
  and\ \citenamefont {Allen}}]{Forbes2010QEP}%
  \BibitemOpen
  \bibfield  {author} {\bibinfo {author} {\bibfnamefont {B.~D.}\ \bibnamefont
  {Forbes}}, \bibinfo {author} {\bibfnamefont {A.~V.}\ \bibnamefont {Martin}},
  \bibinfo {author} {\bibfnamefont {S.~D.}\ \bibnamefont {Findlay}}, \bibinfo
  {author} {\bibfnamefont {A.~J.}\ \bibnamefont {D'Alfonso}},\ and\ \bibinfo
  {author} {\bibfnamefont {L.~J.}\ \bibnamefont {Allen}},\ }\href
  {https://doi.org/10.1103/PhysRevB.82.104103} {\bibfield  {journal} {\bibinfo
  {journal} {Phys. Rev. B}\ }\textbf {\bibinfo {volume} {82}},\ \bibinfo
  {pages} {104103} (\bibinfo {year} {2010})}\BibitemShut {NoStop}%
\bibitem [{Upp(2023)}]{UppASD}%
  \BibitemOpen
  \href@noop {} {\bibinfo {title} {The {U}ppsala atomistic spin dynamics code,
  \texttt{UppASD}}},\ \bibinfo {howpublished}
  {\url{https://github.com/UppASD/UppASD}} (\bibinfo {year} {2023}),\ \bibinfo
  {note} {last accessed 2023-01-30.}\BibitemShut {Stop}%
\bibitem [{Note1()}]{Note1}%
  \BibitemOpen
  \bibinfo {note} {We have used \protect \texttt {UppASD} version
  5.0.}\BibitemShut {Stop}%
\bibitem [{\citenamefont {Ebert}\ \emph {et~al.}(2011)\citenamefont {Ebert},
  \citenamefont {Ködderitzsch},\ and\ \citenamefont {Minár}}]{Ebert2011}%
  \BibitemOpen
  \bibfield  {author} {\bibinfo {author} {\bibfnamefont {H.}~\bibnamefont
  {Ebert}}, \bibinfo {author} {\bibfnamefont {D.}~\bibnamefont
  {Ködderitzsch}},\ and\ \bibinfo {author} {\bibfnamefont {J.}~\bibnamefont
  {Minár}},\ }\href {https://doi.org/10.1088/0034-4885/74/9/096501} {\bibfield
   {journal} {\bibinfo  {journal} {Rep. Prog. Phys.}\ }\textbf {\bibinfo
  {volume} {74}},\ \bibinfo {pages} {096501} (\bibinfo {year}
  {2011})}\BibitemShut {NoStop}%
\bibitem [{lam()}]{lammps_web}%
  \BibitemOpen
  \href@noop {} {\bibinfo {title} {The {L}arge-scale {A}tomic/{M}olecular
  {M}assively {P}arallel {S}imulator ({LAMMPS})}},\ \bibinfo {howpublished}
  {\url{https://www.lammps.org}}\BibitemShut {NoStop}%
\bibitem [{\citenamefont {Thompson}\ \emph {et~al.}(2022)\citenamefont
  {Thompson}, \citenamefont {Aktulga}, \citenamefont {Berger}, \citenamefont
  {Bolintineanu}, \citenamefont {Brown}, \citenamefont {Crozier}, \citenamefont
  {in~'t Veld}, \citenamefont {Kohlmeyer}, \citenamefont {Moore}, \citenamefont
  {Nguyen}, \citenamefont {Shan}, \citenamefont {Stevens}, \citenamefont
  {Tranchida}, \citenamefont {Trott},\ and\ \citenamefont
  {Plimpton}}]{LAMMPS_paper_2022}%
  \BibitemOpen
  \bibfield  {author} {\bibinfo {author} {\bibfnamefont {A.~P.}\ \bibnamefont
  {Thompson}}, \bibinfo {author} {\bibfnamefont {H.~M.}\ \bibnamefont
  {Aktulga}}, \bibinfo {author} {\bibfnamefont {R.}~\bibnamefont {Berger}},
  \bibinfo {author} {\bibfnamefont {D.~S.}\ \bibnamefont {Bolintineanu}},
  \bibinfo {author} {\bibfnamefont {W.~M.}\ \bibnamefont {Brown}}, \bibinfo
  {author} {\bibfnamefont {P.~S.}\ \bibnamefont {Crozier}}, \bibinfo {author}
  {\bibfnamefont {P.~J.}\ \bibnamefont {in~'t Veld}}, \bibinfo {author}
  {\bibfnamefont {A.}~\bibnamefont {Kohlmeyer}}, \bibinfo {author}
  {\bibfnamefont {S.~G.}\ \bibnamefont {Moore}}, \bibinfo {author}
  {\bibfnamefont {T.~D.}\ \bibnamefont {Nguyen}}, \bibinfo {author}
  {\bibfnamefont {R.}~\bibnamefont {Shan}}, \bibinfo {author} {\bibfnamefont
  {M.~J.}\ \bibnamefont {Stevens}}, \bibinfo {author} {\bibfnamefont
  {J.}~\bibnamefont {Tranchida}}, \bibinfo {author} {\bibfnamefont
  {C.}~\bibnamefont {Trott}},\ and\ \bibinfo {author} {\bibfnamefont {S.~J.}\
  \bibnamefont {Plimpton}},\ }\href {https://doi.org/10.1016/j.cpc.2021.108171}
  {\bibfield  {journal} {\bibinfo  {journal} {Comp. Phys. Comm.}\ }\textbf
  {\bibinfo {volume} {271}},\ \bibinfo {pages} {108171} (\bibinfo {year}
  {2022})}\BibitemShut {NoStop}%
\bibitem [{Note2()}]{Note2}%
  \BibitemOpen
  \bibinfo {note} {We have used \protect \texttt {LAMMPS} version 23 Jun 2022
  -- Update 1.}\BibitemShut {Stop}%
\bibitem [{\citenamefont {Mendelev}\ \emph {et~al.}(2003)\citenamefont
  {Mendelev}, \citenamefont {Han}, \citenamefont {Srolovitz}, \citenamefont
  {Ackland}, \citenamefont {Sun},\ and\ \citenamefont
  {Asta}}]{Mendelev2003PhilMag}%
  \BibitemOpen
  \bibfield  {author} {\bibinfo {author} {\bibfnamefont {M.~I.}\ \bibnamefont
  {Mendelev}}, \bibinfo {author} {\bibfnamefont {S.}~\bibnamefont {Han}},
  \bibinfo {author} {\bibfnamefont {D.~J.}\ \bibnamefont {Srolovitz}}, \bibinfo
  {author} {\bibfnamefont {G.~J.}\ \bibnamefont {Ackland}}, \bibinfo {author}
  {\bibfnamefont {D.~Y.}\ \bibnamefont {Sun}},\ and\ \bibinfo {author}
  {\bibfnamefont {M.}~\bibnamefont {Asta}},\ }\href
  {https://doi.org/10.1080/14786430310001613264} {\bibfield  {journal}
  {\bibinfo  {journal} {Philos. Mag.}\ }\textbf {\bibinfo {volume} {83}},\
  \bibinfo {pages} {3977} (\bibinfo {year} {2003})}\BibitemShut {NoStop}%
\bibitem [{Note3()}]{Note3}%
  \BibitemOpen
  \bibinfo {note} {The smallest achievable size of the employed electron probe
  on the specimen is $d_0=1.5311$~nm, corresponding roughly to a linear
  extension of 5.3 bcc Fe unit cells. This value can be computed from the
  diffraction limit $d_0=0.61\lambda /\alpha _0$~\cite {pennycook2011scanning},
  using the probe's convergence semiangle $\alpha _0=1$~mrad and its 200~kV
  de-Broglie wavelength $\lambda =0.00251$~nm~\cite
  {kirkland2010}.}\BibitemShut {Stop}%
\bibitem [{\citenamefont {Lyon}\ and\ \citenamefont {Rusz}(2021)}]{Lyon2021AC}%
  \BibitemOpen
  \bibfield  {author} {\bibinfo {author} {\bibfnamefont {K.}~\bibnamefont
  {Lyon}}\ and\ \bibinfo {author} {\bibfnamefont {J.}~\bibnamefont {Rusz}},\
  }\href {https://doi.org/10.1107/S2053273321008792} {\bibfield  {journal}
  {\bibinfo  {journal} {Acta Cryst. A}\ }\textbf {\bibinfo {volume} {77}},\
  \bibinfo {pages} {509} (\bibinfo {year} {2021})}\BibitemShut {NoStop}%
\bibitem [{\citenamefont {Dudarev}\ \emph {et~al.}(1995)\citenamefont
  {Dudarev}, \citenamefont {Peng},\ and\ \citenamefont
  {Whelan}}]{DUDAREV199586}%
  \BibitemOpen
  \bibfield  {author} {\bibinfo {author} {\bibfnamefont {S.}~\bibnamefont
  {Dudarev}}, \bibinfo {author} {\bibfnamefont {L.-M.}\ \bibnamefont {Peng}},\
  and\ \bibinfo {author} {\bibfnamefont {M.}~\bibnamefont {Whelan}},\ }\href
  {https://doi.org/https://doi.org/10.1016/0039-6028(95)00464-5} {\bibfield
  {journal} {\bibinfo  {journal} {Surf. Sci.}\ }\textbf {\bibinfo {volume}
  {330}},\ \bibinfo {pages} {86} (\bibinfo {year} {1995})}\BibitemShut
  {NoStop}%
\bibitem [{\citenamefont {Peng}\ \emph {et~al.}(1996)\citenamefont {Peng},
  \citenamefont {Ren}, \citenamefont {Dudarev},\ and\ \citenamefont
  {Whelan}}]{Peng1996}%
  \BibitemOpen
  \bibfield  {author} {\bibinfo {author} {\bibfnamefont {L.-M.}\ \bibnamefont
  {Peng}}, \bibinfo {author} {\bibfnamefont {G.}~\bibnamefont {Ren}}, \bibinfo
  {author} {\bibfnamefont {S.~L.}\ \bibnamefont {Dudarev}},\ and\ \bibinfo
  {author} {\bibfnamefont {M.~J.}\ \bibnamefont {Whelan}},\ }\href
  {https://doi.org/10.1107/S0108767395014371} {\bibfield  {journal} {\bibinfo
  {journal} {Acta Cryst. A}\ }\textbf {\bibinfo {volume} {52}},\ \bibinfo
  {pages} {257} (\bibinfo {year} {1996})}\BibitemShut {NoStop}%
\bibitem [{\citenamefont {Carter}\ and\ \citenamefont
  {Williams}(2016)}]{carter2016transmission}%
  \BibitemOpen
  \bibfield  {author} {\bibinfo {author} {\bibfnamefont {C.~B.}\ \bibnamefont
  {Carter}}\ and\ \bibinfo {author} {\bibfnamefont {D.~B.}\ \bibnamefont
  {Williams}},\ }\href
  {https://doi.org/https://doi.org/10.1007/978-3-319-26651-0} {\emph {\bibinfo
  {title} {Transmission Electron Microscopy: Diffraction, Imaging, and
  Spectrometry}}}\ (\bibinfo  {publisher} {Springer},\ \bibinfo {year}
  {2016})\BibitemShut {NoStop}%
\bibitem [{\citenamefont {Rose}(1973)}]{Rose1973}%
  \BibitemOpen
  \bibfield  {author} {\bibinfo {author} {\bibfnamefont {A.}~\bibnamefont
  {Rose}},\ }\href {https://doi.org/https://doi.org/10.1007/978-1-4684-2037-1}
  {\emph {\bibinfo {title} {Vision: Human and Electronic}}}\ (\bibinfo
  {publisher} {Springer},\ \bibinfo {year} {1973})\BibitemShut {NoStop}%
\bibitem [{\citenamefont {Houdellier}\ \emph {et~al.}(2015)\citenamefont
  {Houdellier}, \citenamefont {{de Knoop}}, \citenamefont {Gatel},
  \citenamefont {Masseboeuf}, \citenamefont {Mamishin}, \citenamefont
  {Taniguchi}, \citenamefont {Delmas}, \citenamefont {Monthioux}, \citenamefont
  {Hÿtch},\ and\ \citenamefont {Snoeck}}]{Houdellier2015107}%
  \BibitemOpen
  \bibfield  {author} {\bibinfo {author} {\bibfnamefont {F.}~\bibnamefont
  {Houdellier}}, \bibinfo {author} {\bibfnamefont {L.}~\bibnamefont {{de
  Knoop}}}, \bibinfo {author} {\bibfnamefont {C.}~\bibnamefont {Gatel}},
  \bibinfo {author} {\bibfnamefont {A.}~\bibnamefont {Masseboeuf}}, \bibinfo
  {author} {\bibfnamefont {S.}~\bibnamefont {Mamishin}}, \bibinfo {author}
  {\bibfnamefont {Y.}~\bibnamefont {Taniguchi}}, \bibinfo {author}
  {\bibfnamefont {M.}~\bibnamefont {Delmas}}, \bibinfo {author} {\bibfnamefont
  {M.}~\bibnamefont {Monthioux}}, \bibinfo {author} {\bibfnamefont
  {M.}~\bibnamefont {Hÿtch}},\ and\ \bibinfo {author} {\bibfnamefont
  {E.}~\bibnamefont {Snoeck}},\ }\href
  {https://doi.org/https://doi.org/10.1016/j.ultramic.2014.11.021} {\bibfield
  {journal} {\bibinfo  {journal} {Ultramicroscopy}\ }\textbf {\bibinfo {volume}
  {151}},\ \bibinfo {pages} {107} (\bibinfo {year} {2015})}\BibitemShut
  {NoStop}%
\bibitem [{\citenamefont {Shao}\ \emph {et~al.}(2018)\citenamefont {Shao},
  \citenamefont {Srinivasan}, \citenamefont {Ang},\ and\ \citenamefont
  {Khursheed}}]{shao2018high}%
  \BibitemOpen
  \bibfield  {author} {\bibinfo {author} {\bibfnamefont {X.}~\bibnamefont
  {Shao}}, \bibinfo {author} {\bibfnamefont {A.}~\bibnamefont {Srinivasan}},
  \bibinfo {author} {\bibfnamefont {W.~K.}\ \bibnamefont {Ang}},\ and\ \bibinfo
  {author} {\bibfnamefont {A.}~\bibnamefont {Khursheed}},\ }\href
  {https://doi.org/https://doi.org/10.1038/s41467-018-03721-y} {\bibfield
  {journal} {\bibinfo  {journal} {Nat. Commun.}\ }\textbf {\bibinfo {volume}
  {9}},\ \bibinfo {pages} {1288} (\bibinfo {year} {2018})}\BibitemShut
  {NoStop}%
\bibitem [{\citenamefont {Konings}\ and\ \citenamefont
  {Bischoff}(2020)}]{konings2020}%
  \BibitemOpen
  \bibfield  {author} {\bibinfo {author} {\bibfnamefont {S.}~\bibnamefont
  {Konings}}\ and\ \bibinfo {author} {\bibfnamefont {M.}~\bibnamefont
  {Bischoff}},\ }\href {https://doi.org/10.1017/S143192762001510X} {\bibfield
  {journal} {\bibinfo  {journal} {Microsc. Microanal.}\ }\textbf {\bibinfo
  {volume} {26}},\ \bibinfo {pages} {566–567} (\bibinfo {year}
  {2020})}\BibitemShut {NoStop}%
\bibitem [{\citenamefont {Egerton}(2011)}]{egerton2011electron}%
  \BibitemOpen
  \bibfield  {author} {\bibinfo {author} {\bibfnamefont {R.~F.}\ \bibnamefont
  {Egerton}},\ }\href
  {https://doi.org/https://doi.org/10.1007/978-1-4419-9583-4} {\emph {\bibinfo
  {title} {Electron Energy-loss Spectroscopy in the Electron Microscope}}}\
  (\bibinfo  {publisher} {Springer Science \& Business Media},\ \bibinfo {year}
  {2011})\BibitemShut {NoStop}%
\bibitem [{\citenamefont {Zeltmann}\ \emph {et~al.}(2020)\citenamefont
  {Zeltmann}, \citenamefont {Müller}, \citenamefont {Bustillo}, \citenamefont
  {Savitzky}, \citenamefont {Hughes}, \citenamefont {Minor},\ and\
  \citenamefont {Ophus}}]{zeltmann}%
  \BibitemOpen
  \bibfield  {author} {\bibinfo {author} {\bibfnamefont {S.~E.}\ \bibnamefont
  {Zeltmann}}, \bibinfo {author} {\bibfnamefont {A.}~\bibnamefont {Müller}},
  \bibinfo {author} {\bibfnamefont {K.~C.}\ \bibnamefont {Bustillo}}, \bibinfo
  {author} {\bibfnamefont {B.}~\bibnamefont {Savitzky}}, \bibinfo {author}
  {\bibfnamefont {L.}~\bibnamefont {Hughes}}, \bibinfo {author} {\bibfnamefont
  {A.~M.}\ \bibnamefont {Minor}},\ and\ \bibinfo {author} {\bibfnamefont
  {C.}~\bibnamefont {Ophus}},\ }\href
  {https://doi.org/https://doi.org/10.1016/j.ultramic.2019.112890} {\bibfield
  {journal} {\bibinfo  {journal} {Ultramicroscopy}\ }\textbf {\bibinfo {volume}
  {209}},\ \bibinfo {pages} {112890} (\bibinfo {year} {2020})}\BibitemShut
  {NoStop}%
\bibitem [{Note4()}]{Note4}%
  \BibitemOpen
  \bibinfo {note} {In Ref.~\cite {DUDAREV199586}, the parameters $a_n^\protect
  \text {(abs)}$ and $b_n^\protect \text {(abs)}$ are called $a_n^\protect
  \text {(TDS)}$ and $b_n^\protect \text {(TDS)}$, respectively.}\BibitemShut
  {Stop}%
\bibitem [{\citenamefont {Zeiger}\ \emph {et~al.}(2023)\citenamefont {Zeiger},
  \citenamefont {Barthel}, \citenamefont {Allen},\ and\ \citenamefont
  {Rusz}}]{zeiger2023lessons}%
  \BibitemOpen
  \bibfield  {author} {\bibinfo {author} {\bibfnamefont {P.~M.}\ \bibnamefont
  {Zeiger}}, \bibinfo {author} {\bibfnamefont {J.}~\bibnamefont {Barthel}},
  \bibinfo {author} {\bibfnamefont {L.~J.}\ \bibnamefont {Allen}},\ and\
  \bibinfo {author} {\bibfnamefont {J.}~\bibnamefont {Rusz}},\ }\href
  {https://doi.org/10.1103/PhysRevB.108.094309} {\bibfield  {journal} {\bibinfo
   {journal} {Phys. Rev. B}\ }\textbf {\bibinfo {volume} {108}},\ \bibinfo
  {pages} {094309} (\bibinfo {year} {2023})}\BibitemShut {NoStop}%
\bibitem [{\citenamefont {Malica}\ and\ \citenamefont
  {Dal~Corso}(2019)}]{Malica:ae5064}%
  \BibitemOpen
  \bibfield  {author} {\bibinfo {author} {\bibfnamefont {C.}~\bibnamefont
  {Malica}}\ and\ \bibinfo {author} {\bibfnamefont {A.}~\bibnamefont
  {Dal~Corso}},\ }\href {https://doi.org/10.1107/S205327331900514X} {\bibfield
  {journal} {\bibinfo  {journal} {Acta Cryst. A}\ }\textbf {\bibinfo {volume}
  {75}},\ \bibinfo {pages} {624} (\bibinfo {year} {2019})}\BibitemShut
  {NoStop}%
\bibitem [{\citenamefont {Martin}\ \emph {et~al.}(2009)\citenamefont {Martin},
  \citenamefont {Findlay},\ and\ \citenamefont {Allen}}]{AllenPhysRevB}%
  \BibitemOpen
  \bibfield  {author} {\bibinfo {author} {\bibfnamefont {A.~V.}\ \bibnamefont
  {Martin}}, \bibinfo {author} {\bibfnamefont {S.~D.}\ \bibnamefont
  {Findlay}},\ and\ \bibinfo {author} {\bibfnamefont {L.~J.}\ \bibnamefont
  {Allen}},\ }\href {https://doi.org/10.1103/PhysRevB.80.024308} {\bibfield
  {journal} {\bibinfo  {journal} {Phys. Rev. B}\ }\textbf {\bibinfo {volume}
  {80}},\ \bibinfo {pages} {024308} (\bibinfo {year} {2009})}\BibitemShut
  {NoStop}%
\bibitem [{\citenamefont {Findlay}\ \emph {et~al.}(2008)\citenamefont
  {Findlay}, \citenamefont {Oxley},\ and\ \citenamefont {Allen}}]{AllenMandM}%
  \BibitemOpen
  \bibfield  {author} {\bibinfo {author} {\bibfnamefont {S.~D.}\ \bibnamefont
  {Findlay}}, \bibinfo {author} {\bibfnamefont {M.~P.}\ \bibnamefont {Oxley}},\
  and\ \bibinfo {author} {\bibfnamefont {L.~J.}\ \bibnamefont {Allen}},\ }\href
  {https://doi.org/10.1017/S1431927608080112} {\bibfield  {journal} {\bibinfo
  {journal} {Microsc. Microanal.}\ }\textbf {\bibinfo {volume} {14}},\ \bibinfo
  {pages} {48–59} (\bibinfo {year} {2008})}\BibitemShut {NoStop}%
\bibitem [{\citenamefont {Pennycook}\ and\ \citenamefont
  {Nellist}(2011)}]{pennycook2011scanning}%
  \BibitemOpen
  \bibfield  {author} {\bibinfo {author} {\bibfnamefont {S.~J.}\ \bibnamefont
  {Pennycook}}\ and\ \bibinfo {author} {\bibfnamefont {P.~D.}\ \bibnamefont
  {Nellist}},\ }\href
  {https://doi.org/https://doi.org/10.1007/978-1-4419-7200-2} {\emph {\bibinfo
  {title} {Scanning Transmission Electron Microscopy: Imaging and Analysis}}}\
  (\bibinfo  {publisher} {Springer Science \& Business Media},\ \bibinfo {year}
  {2011})\BibitemShut {NoStop}%
\end{thebibliography}%


\end{document}